\documentclass[twocolumn,prb,showpacs,preprintnumbers,amsmath,amssymb]{revtex4}

\usepackage{graphicx}
\usepackage{dcolumn}
\usepackage{bm}
\usepackage{epsf}

\usepackage{multirow}

\begin{document}
\newcommand{\cm}{\ \mbox{cm}^{-1} }

\title{Surface Loving  and Surface Avoiding modes}

\author{Nicolas Combe$^1$}
\email{combe@cemes.fr}
\author{Jean Roch Huntzinger$^2$}
\email{jean-roch.huntzinger@univ-montp2.fr}
\author{Joseph Morillo$^1$}
\email{morillo@cemes.fr}
\affiliation{$^1$ Centre d'Elaboration de Mat\'eriaux et d'Etudes
Structurales, CNRS UPR 8011-Universit\'e Paul Sabatier, 29 rue J. Marvig, BP 94347, 31055 Toulouse cedex
4, France}

\affiliation{$^2$Groupe d'Etude des Semiconducteurs,  UMR 5650 CNRS-Universit\'e Montpellier II
Bt 21, CC074,
Place Eug\`ene Bataillon,
34095 Montpellier Cedex 05
France}

\providecommand{\abs}[1]{\ensuremath{\left | #1 \right |}}
\providecommand{\bloc}[1]{\ensuremath{\left ( #1 \right )}}
\providecommand{\bloca}[1]{\ensuremath{\left \{ #1 \right \}}}
\providecommand{\blocc}[1]{\ensuremath{\left [ #1 \right ]}}
\providecommand{\mean}[1]{\ensuremath{\left \langle #1 \right \rangle}}
\providecommand{\modulo}[1]{\ensuremath{\,[#1]}}
\providecommand{\ket}[1]{\ensuremath{\left | #1 \right \rangle}}
\providecommand{\bra}[1]{\ensuremath{\left \langle #1 \right |}}
\providecommand{\braket}[2]{\ensuremath{\left \langle   #1  |  #2 \right \rangle}}
\date{\today}
\def\nbN{\ensuremath{\mathrm{I\!N}}} 
\providecommand{\conj}[1]{\ensuremath{\overline{#1}}}

\providecommand{\cmoy}{\ensuremath{c_0}}
\providecommand{\utot}{\ensuremath{\vec{u}}}
\providecommand{\urms}{\ensuremath{u_\text{rms}}}
\providecommand{\nuhalfbelow}{0.5807}
\providecommand{\nuhalfabove}{0.6182}
\providecommand{\sayparam}{Parameters corresponding to the SL described in section~\ref{diag_disp} have been used.}


\begin{abstract}
 
We theoretically study the propagation of sound waves in GaAs/AlAs superlattices focussing on periodic modes in the vicinity of the band gaps. Based on analytical and numerical calculations,  we show that these modes are the product of a quickly oscillating function times a slowly varying envelope function. We carefully study the phase of the envelope function compared to the surface of a semi-infinite superlattice. Especially, the dephasing of the superlattice compared to its surface is a key parameter.  We exhibit two kind of modes: Surface Avoiding and Surface Loving  Modes whose envelope functions have their minima and respectively maxima in the vicinity of the surface.  We finally consider the observability of such modes. While Surface avoiding modes have experimentally been observed (Phys. Rev. Lett. 97, 1224301 (2006)), we show that Surface Loving Modes are likely to be observable and we discuss the achievement of such experiments. 
The proposed approach could be easily transposed to other types of wave propagation in unidimensional semi-infinite periodic structures as photonic Bragg mirror.
\end{abstract}


\maketitle 
\section{Introduction}

Propagation of waves in periodic structures have been studied for decades~\cite{brillouin}. Indeed, the periodicity leads to band gaps where propagation is forbidden. This property is very general and could be observed in many different fields. Electrons in crystals experience the periodic atomic potential leading to electronic band gaps~\cite{ashcroft}.  Electromagnetic waves cannot propagate in a spatial periodic dielectric constant system if their frequencies fall in the band gap:  Photonic crystals~\cite{yablonovitch,john} are based on this statement.  Elastic waves in solids could also endure a periodic Young modulus and/or a periodic density. Studies on such systems are currently numerous~\cite{winter,kent2006,tamura:165306} and lead to a new field of physics namely  Phononics~\cite{gorishny,meseguer}.

In periodic structures, two kinds of vibrations can be pointed out: 1) extended eigenmodes that propagate through the system;  their  frequencies fall out of the gaps and they satisfy the Floquet-Bloch theorem;  2) localized modes  that can be found around isolated defects including surfaces~\cite{tamm}(surface modes belong to this category). Due to their localized character, these modes are not so sensitive to the periodicity of the underlying Crystal and thus their frequency can take any value, in particular within the forbidden band gaps.  

Extended acoustical vibrations in superlattices (SL) have been described quantitatively in the frame of the elastic model in the middle of the century~\cite{rytov}.
The superlattice, with its long period compared to the underlying crystals periods, results in a much shorter reciprocal Brillouin zone and hence a multiple folding of the initial acoustical branches.   Experimentally, observations of these folded acoustical vibrations have first been performed using Raman scattering experiments for which selection rules are now well understood~\cite{colvard,jusserand87}.  Since the nineties, time-domain optical experiments have investigated folded acoustical vibrations~\cite{bartels,mizoguchi,takeuchi,hudert}. They  exhibit two types of modes: phonons with finite wave vectors ($q \neq  0$) and zone-center modes ($q = 0$) in the reduced Brillouin zone.  

 Recently, Trigo et al.~\cite{trigo} studied both theoretically and experimentally  these zone-center modes in phononic semi-infinite GaAs/AlAs SL yielding two conclusions. First, modes near the Brillouin zone-center observed  by pump-probe experiments  have not rigorously  a zero wave vector, but instead  an almost zero  wave vector $q \approx 0$. Second,  these modes present a slowly varying envelope wave function with an amplitude minimum  in the vicinity of the surface:  they claim these modes ``have a tendency to avoid the boundaries, irrespective of the boundary conditions" and for this reason, refer to them as Surface Avoiding Modes (SAM).
Note that these surface avoiding modes have also been encountered near photonic~\cite{Scalora1996} and acoustic~\cite{Aynaou2005} band gap materials and their nature is very general. 
In this work, we study in detail acoustical modes in the vicinity of Brillouin zone center, focussing on their avoiding character near the GaAs/AlAs SL surface. \\
Our work is  based on analytical studies corroborated by numerical simulations.   Analytical calculations are based on the analogy between the propagation of waves in a periodic structure and the spatial parametric oscillator. 
Since this analogy is not well-known, we devote  Sect.~\ref{Param_oscill}  to prove  the fruitfulness of such analogy to describe waves in periodic structures in the vicinity of gaps. Considering the propagation in an infinite SL of longitudinal modes in the vicinity of band gaps, we show that these modes are the product of a fast oscillating function and a slowly varying envelope function.   We derive the dispersion diagram in the vicinity of band gaps using the parametric oscillator equations. Then considering the case of a semi-infinite SL in Sect.~\ref{sam-slm}, we give the analytical expression of longitudinal waves assuming free boundary conditions and considering the dephasing of the SL compared to its surface. We show the existence of Surface Avoiding and Surface Loving Modes depending if the envelope function is  minimum or maximum near the SL surface. We then discuss in detail the avoiding or loving character of modes near the center of the Brillouin zone. In Sect.~\ref{observation}, we discuss on the electron-phonon coupling of these modes and show both Surface Avoiding and Loving Modes are likely to be  observable using time-domain optical experiments or Raman and Brillouin scattering experiments. Finally, Sect.~\ref{discussion} will be devoted to a discussion on our results.

\section{Parametric oscillator} 
\label{Param_oscill}

\subsection{Parametric oscillator analogy}
\label{analogy}
We consider the propagation of a sound wave in an unidimensional infinite SL made of 2 solid materials: respectively named A and B. 
$d_A$ and $d_B$ are  the width of the layers, and $d= d_A + d_B$ is the period of the SL\@.  
If $\vec{u}$ is the atomic displacement, in each material sound waves are solutions of: 
\begin{equation}
\Delta \utot- \frac{1}{c_i^2}  \frac{\partial^2 \utot}{\partial t^2} = \vec{0}
\label{eq1}
\end{equation}
where $c_i$ is the sound speed in the materials ($ i = {A,B}$).
To determinate sound waves in a SL from Eq.~\eqref{eq1}, one needs to apply  at each interface, the continuities of the displacement field and of forces $d \vec{F}=\bar{\bar{\sigma}} d\vec{S}$ acting on each elementary interface $d\vec{S}$,  where $\bar{\bar{\sigma}}$ is the stress tensor.

In a SL, the displacements can be determined exactly by the transfer matrix method~\cite{he}. 
Though very powerful, and clearly grounded, this method suffers from its numerical nature. It is not easy to get analytical expressions for finite size superlattices.

 In this work, we will use the analogy between the propagation of sound waves in periodic media and the parametric oscillator. As it will be shown, this analogy involves severals approximations we will justify. However, it has two benefits.
First, it yields analytical expressions of the displacements in closed form, with relatively simple calculations. 
Second, and especially, it allows to point out and to understand the main physical phenomena near the Bragg reflexions in the SL\@.  
We stress that the present treatment is complementary to the transfer matrix method and does not pretend to challenge it for quantitative analysis. 

 Finally, one must note that  the present method could be applied to the propagation of light in photonic crystals or electrons in crystals~\cite{kushwaha,Combe2008} since their wave propagation equations are similar.

 From now on, we consider a harmonic, plane, longitudinal sound wave of angular frequency $\omega$:  $\utot(z,t)=u(z) e^{-i\omega t} \vec{e_z}$
along the  $z$-axis, perpendicular to the SL layers. Continuity of  forces acting on each interface imposes the continuity of $C_{11}(z) \frac{\partial u(z)}{dz}$ where $C_{11}$ is the stiffness constant, for a longitudinal sound wave propagating along $z$. In the following, we will focus on GaAs/AlAs superlattices, because they are the most experimentally used superlattices.   Stiffness constant contrast between GaAs and AlAs is very weak: $\frac{\Delta C_{11}}{C_{11}} \approx 0.014$  so that the continuity of $C_{11}(z) \frac{\partial u(z)}{dz}$ at each interface reduces to the continuity of $\frac{\partial u(z)}{dz}$.  In such a case, the determination of sound waves in the SL can be directly achieved  by the resolution of Eq.~\eqref{eq1}  considering  only the variation of the sound speed  since $u(z)$ becomes a $\mathcal{C}^2$ function~\footnote{An extension of the following calculations to SL with high stiffness constant contrast could be achieved, we expect the same qualitative conclusions.}.  Consequently, since the sound speed in a GaAs/AlAs SL is a periodic function of $z$, we may expand $\frac{1}{c^2}$ as a Fourier series $\frac{1}{c^2}= a_0 + \sum^{\infty}_{m=1}
a_m \cos(\frac{2 \pi m z}{d}) + b_m \sin(\frac{2 \pi m z}{d})$, so that sound waves in GaAs/AlAs superlattices are solutions of: 
\begin{equation} 
\frac{d^2u}{dz^2} + \omega^2 a_0 \left[1 + \sum^{\infty}_{m=1} \frac{a_m}{a_0} 
\cos(\frac{2 \pi m z}{d}) + \frac{b_m}{a_0} \sin(\frac{2 \pi m z}{d})  \right]  u =0 \label{eq2}
\end{equation}

We now introduce $G =\frac{2 \pi}{d}$ the primitive  vector of the reciprocal lattice, $c_m= \frac{a_m}{a_0}$ and $d_m = \frac{b_m}{a_0}$ the reduced amplitudes of the $m$th harmonic, and  $k_0(\omega) = \omega \sqrt{a_0}$  the zero order wave number i.e.\ the wave number a wave would have in the absence of any periodic modulation of the sound speed.  $k_0(\omega) = \omega \sqrt{a_0}$ is the usual dispersion relation for acoustic phonons in an homogeneous medium with mean sound speed $c_0=\frac{1}{\sqrt{a_0}}$. Eq~\eqref{eq2} then reads  like a parametric oscillator equation in spatial coordinates: 
 \begin{equation} 
\frac{d^2u}{dz^2} + k_0^2(\omega)  \left[1 + \sum^{\infty}_{m=1} c_m \cos(m G z) + d_m \sin(m G z) \right] u=0
\label{eq_gene}
\end{equation} 

\subsection{Single Harmonic Approximation} 
\label{import_harm}
\begin{figure}
\centerline{\epsfxsize=8cm  \epsfbox{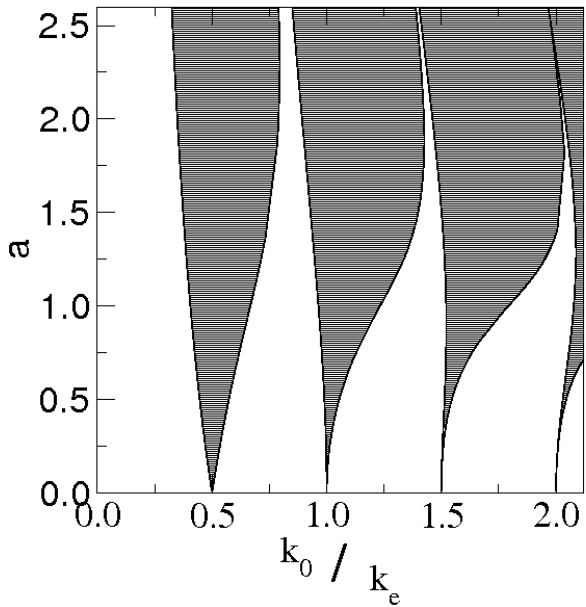} }
\caption{Phase diagram of the Mathieu equation. Parameters inducing non periodic solutions are in dashed regions (``parametric resonances" or ``band gaps").}
\label{fig1}
\end{figure}

Eq.~\eqref{eq_gene} includes all harmonics of the inverse square sound velocity. In order to understand the effect induced by the different harmonics, let us first consider the effect of an isolated harmonic. Eq.~\eqref{eq_gene}, in this case, reduces to a Mathieu equation~\cite{Abramowitz} of the form:
\begin{equation}
\frac{d^2 u}{dz^2} + k_0^2 \left[ 1 + a \cos (k_e z) \right]  u =0 \label{eq_mathieu}
\end{equation}
$k_e$ is the wave number of the periodic excitation. 
General solutions of Eq.~\eqref{eq_mathieu} are  periodic except in some regions of the plane $(k_0/k_e,a)$: they are then the product of an oscillating function with a linear combination of an increasing and a decreasing exponential.  These non periodic solutions correspond to the resonances of the parametric oscillator or band gaps of the SL\@.
Fig.~\ref{fig1} presents the stability  diagram of  Eq.~\eqref{eq_mathieu} limited to the first four resonances as a function of  $k_0/k_e$ and $a$: regions of non-periodic solutions are dashed. This diagram has been calculated numerically using the resolvent of Eq.~\eqref{eq_mathieu}  and the Liouville and Floquet-Bloch theorems~\cite{Abramowitz}. 

Regions of non periodic solutions form in the plane  $(k_0/k_e,a)$ bands that, when $a$ tends to 0, converge  to points $\frac{k_0}{k_e} = \frac{n}{2}$ with $n \in \nbN^*$.   i.e.\ wave numbers $k_e = 2k_0, k_0, \frac{2}{3} k_0, \frac{k_0}{2}$, \dots etc.    
Moreover, for $a \ll 1$, it can be shown\cite{landau_meca} that the band width of the $n$th resonance is a decreasing function of  $n$ proportional to $a^n$. As a result,  the larger resonance is obtained for $n=1$ i.e.\ for $k_e = 2k_0$, the most famous resonance of the parametric oscillator, corresponding to an excitation twice faster than the zero order wave number.\\

\begin{figure}
\centerline{\epsfxsize=8cm  \epsfbox{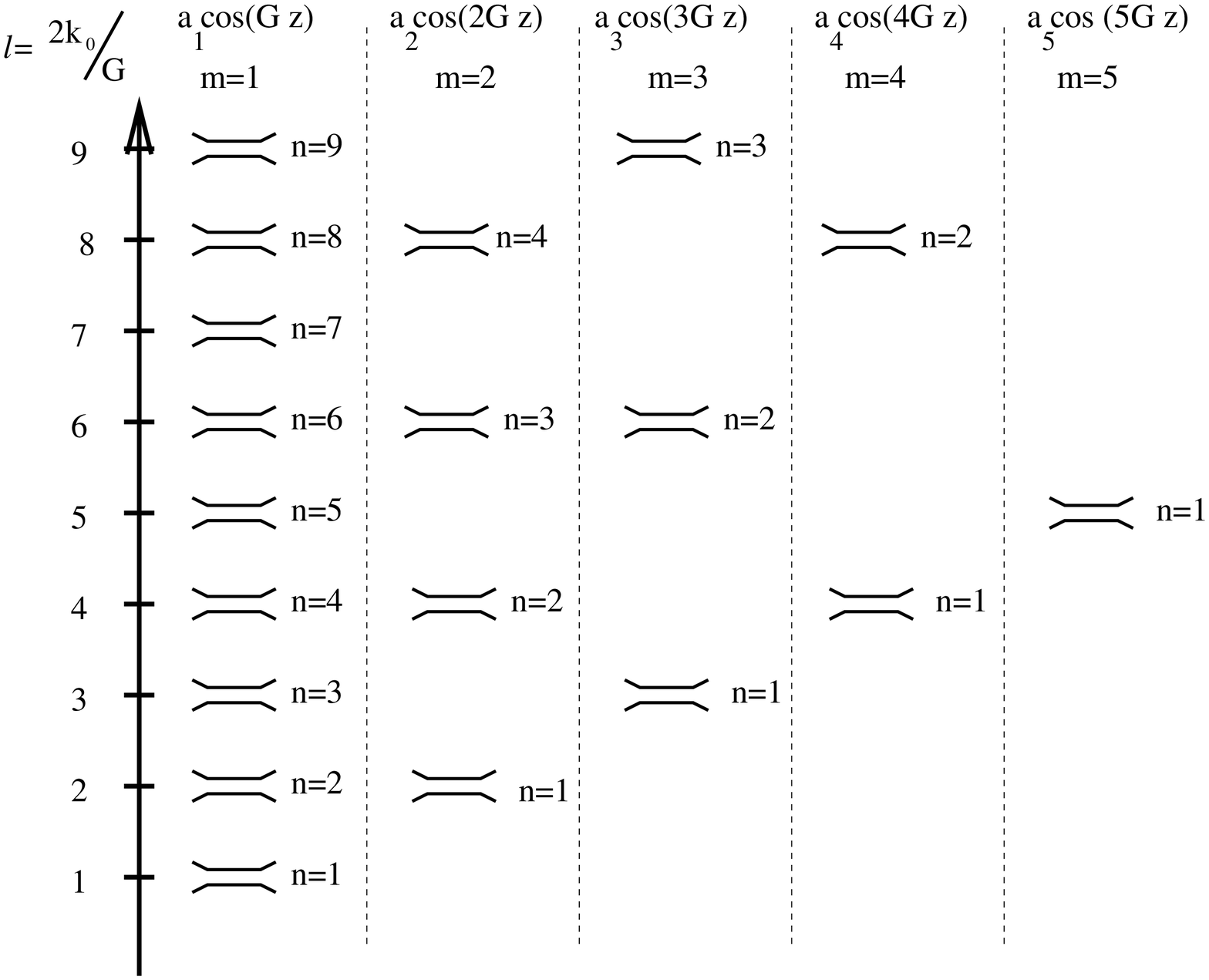} }
\caption{Schematic illustration of band gaps produced by each harmonic.}
\label{fig2}
\end{figure}
Hence, going back to Eq.~\eqref{eq_gene},  varying the value of $k_0$, each harmonic $m$ of the excitation at wave number $m G$, if treated independently,  is expected  to create band gaps around  $\frac{k_0}{m G} = \frac{n}{2}$. Thus, band gaps occur for  $\frac{2 k_0}{G} = nm = l \in \nbN^*$.  Fig.~\ref{fig2}  schematically presents the band gaps independently created by each of the first fifth harmonics.
 Excepted the first  band gap ($l=1$) due exclusively to the first harmonic ($m=1$), all other band gaps result from at least two harmonics. 
However,  as a first approximation, since the larger resonance due to the $m$th harmonic is for $n=1$ i.e.\ around $\frac{k_0}{mG} = \frac{1}{2}$, we consider that the $l$th band gap of Eq.~\eqref{eq_gene} is essentially  due to the first resonance ($n=1$) of the $l$th harmonic ($m=l$). We call this approximation the Single Harmonic Approximation (SHA). This approximation is equivalent to treating the modulation of the sound speed as a first order perturbation: $\abs{c_m} \ll1$ and $\abs{d_m} \ll 1$ $\forall m \in \nbN^*$.  
 Our main motivation for the SHA  is the possibility to derive analytical calculations and the possibility to produce good analytical approximations of  solutions of Eq.~\eqref{eq_gene} around a given gap. Of course, the validity of  SHA will be discussed below in the considered case.    
Note that values of  $k_0$ for which  there are no periodic solution,  correspond to a range of frequencies $\omega$ which is the usual band gap of the SL. Indeed, $k_0$ has been defined by $k_0= \omega \sqrt{a_0}$ in Sect.~\ref{analogy}.     

\subsection{Infinite medium: band gap}  
\label{bandgap}

In order to test the validity of SHA, we  solve Eq.~\eqref{eq_gene} near the resonance~\footnote{A similar study could be done in the vicinity of any band gap as soon as the SHA is checked.} at $\omega_{BG}^{(2)}$ defined by $k_0(\omega_{BG}^{(2)})=G$ i.e. the first resonance due to several harmonics ($m=1$ and  $m=2$). The expression of $\omega_{BG}^{(2)}$ reads: $\omega_{BG}^{(2)} = \frac{G}{\sqrt{a_0}}$. The superscript ``$(2)$'' in $\omega_{BG}^{(2)}$ refers to the second band gap. 
 Using the SHA, this band gap is only produced by the harmonic at $2G$ in the Fourier series of $\frac{1}{c^2}$.  We thus study a reduced equation taking into account only that harmonic: 
 \begin{equation} 
\frac{d^2 u}{dz^2} + k_0^2  \left[ 1 +  c_2 \cos(2 G z) +  d_2 \sin(2 G z) \right] u=0
\label{eq_oscill}
\end{equation}
Where we have simplified the notation of $k_0(\omega)$ using $k_0$. 
The SHA will be checked a posteriori. Switching to complex number representation and introducing $p_2=\frac{ c_2 - i d_2}{2}$, Eq.~\eqref{eq_oscill} reads: 
\begin{equation}
\frac{d^2u}{dz^2} + k_0^2 (1 +  p_2  e^{i2  G z } + \bar{p}_2 e^{-  i2  G z }   ) u = 0, \label{eq_oscill_complx}
\end{equation}
where $\bar{p}_2$ denotes the complex conjugate of $p_2$.
Similarly to the resolution of the parametric oscillator equation by Landau and Lifchitz~\cite{landau_meca},  
we change the coordinate system  to $A(z)$ and $B(z)$ defined by: 
\begin{equation}
u(z) =  A(z)  e^{iG z} + B(z)  e^{-iG z} 
\label{eq_u}
\end{equation}
where $A(z)$ and $B(z)$ are slowly varying functions.
 Introducing  Eq.~\eqref{eq_u} in Eq.~\eqref{eq_oscill_complx} and neglecting fast oscillating terms (at wavevector 3 $G$) as well as the second derivatives of $A(z)$ and $B(z)$, the  identification of coefficients of plane waves at $G$ and $-G$ reads:
 \begin{equation}
 \left[\begin{array}{c}
 \frac{dA}{dz}\\
 \frac{dB}{dz} 
\end{array} \right]  
 =
 \left[\begin{array}{cc}
 -\frac{k_0^2 -G^2 }{2 i G}  &   -\frac{p_2 k_0^2 }{2i G} \\
 \frac{\bar{p}_2 k_0^2 }{2 i G}   & \frac{k_0^2 -G^2 }{2 i G} \\
\end{array} \right]  
\left[\begin{array}{c}
A\\
B
\end{array} \right] 
\label{mat_AB}
\end{equation}
whose general solutions are: 
\begin{subequations}
\label{eqAB}
\begin{align} 
A(z) &= A_+ e^{i \kappa z}  + A_- e^{-i \kappa z} \label{eqA}\\
B(z) &= B_+ e^{i \kappa z}  +  B_- e^{-i \kappa z} \label{eqB}
\end{align}
\end{subequations}
where $\kappa$ is given by: 
\begin{eqnarray}
\kappa^2 &=&  \left[ \frac{ (k_0^2 -G^2)^2}{4 G^2} - \frac{ \abs{p_2}^2 k_0^4}{4 G^2} \right] \nonumber\\
& = &  \bloc{\frac{k_0^2}{2 G}}^2 \bloc{  \delta^2  - \abs{p_2}^2 } \label{ka2} \\
\mbox{ and  } \delta&=&\frac{k_0^2 -G^2}{k_0^2} \label{delta}
\end{eqnarray}
$\delta$ characterizes  the detuning between the excitation wave number $G$ and the zero order wave number $k_0$.\\
The system undergoes a bifurcation when  $\kappa^2$ changes its sign as a function of $\omega$: 
\begin{itemize}
\item Either  $\kappa^2<0$. Solutions of Eq.~\eqref{eqAB}  are then hyperbolic: $A(z)$ and $B(z)$ are a linear combination of increasing and decreasing exponentials. And therefore,  no propagation can occur in the SL\@. The condition $\kappa^2<0$ defines the band gap of the SL around $\omega_{BG}^{(2)}$. 

\item Or $\kappa^2>0$. Solutions of Eq.~\eqref{eqAB} are sinusoidal and the general solution $u(z)$  is thus periodic. Note that in this case, the spectrum of $u(z)$ will be composed of a mixing of $G+\kappa$ and $G-\kappa $ plane waves. $u(z)$ is then the product of a quickly oscillating function at $G$, times a slowly varying functions at $\kappa$. Moreover, Eq.~\eqref{ka2} establishes a relation between $\omega$ (through $k_0(\omega)$)  and wave vectors $G \pm \kappa$ i.e.\  the dispersion relation of the SL around the considered gap. 
\end{itemize}

 Note that the band gap only depends on the  squared magnitude of $p_2$: $\abs{p_2}^2=\frac{ c_2^2 + d_2^2}{4}$ i.e.\ on the amplitude of the second harmonics at $2 G$.\\
Besides, we already pointed out the similarities of sound waves propagation in SL and electron waves propagating in crystals~\cite{kushwaha}. 
Usually, when studying the independent electrons in a weak crystalline potential, one applies  the first order perturbation theory to determine the width and position of electronic band gap~\cite{ashcroft}:  the expression of the band gap width is analogous to  Eq.~\eqref{eqA} and could be derived using the parametric oscillator analogy and the SHA~\cite{Combe2008}.   

\subsection{Dispersion diagram: numerical study}
\label{diag_disp}
In this part, we will numerically check our analytical results  of Sect.~\ref{bandgap} and the SHA applying our predictions on a GaAs/AlAs SL with $d_A=d_{GaAs}=5.9$~nm and $d_B=d_{AlAs} = 2.35$~nm. Such a SL has been used in an experimental setup by Trigo at al.~\cite{trigo}.  For these materials, longitudinal  sound speeds are~\cite{he}: 
\begin{subequations}
\begin{align}
c_A=c_\text{GaAs} &= 4726~\text{ms}^{-1} \\
c_B=c_\text{AlAs} &= 5630~\text{ms}^{-1} 
\end{align} 
\end{subequations}
These parameters will be used in all the numerical examples given in this article.

\begin{figure}
\centerline{\epsfxsize=8cm  \epsfbox{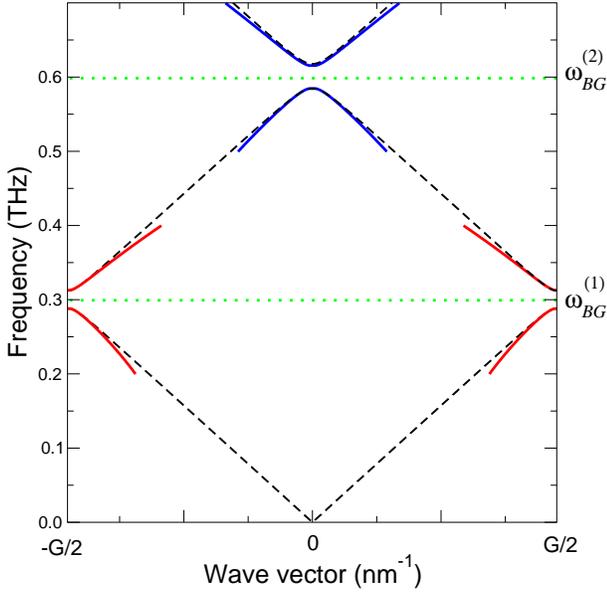} }
\caption{(color online)  Dispersion diagram of a  GaAs/AlAs SL calculated using the transfer matrix method (dashed black) and predictions of the gaps using the parametric oscillator analogy and the SHA: (red) first gap calculated from the first harmonic of the inverse squared sound speed and (blue) second gap calculated from the second harmonic. We report using green horizontal dotted lines the positions of the $l$th band gap  $\omega_{BG}^{(l)} = \frac{lG}{2\sqrt{a_0}}$. \sayparam}
\label{fig3}
\end{figure}

 Fig.~\ref{fig3} compares the dispersion diagram obtained from Eq.~\eqref{ka2} (SHA approx.) and the transfer matrix method~\cite{he}(exact solutions) for this SL\@.  Fig.~\ref{fig3} is restricted  to the two first band gaps respectively due to the harmonics at spatial frequencies  $G$ and $2 G$. We would like to emphasize that in Eq.~\eqref{ka2}, the first band gap is obtained using coefficient $p_1 = \frac{c_1-i d_1}{2}$ (instead of $p_2$) and   the second bang gap using only $p_2 =\frac{ c_2 - i d_2}{2}$ whereas the transfer matrix method considers all harmonics.\\
According to Fig.~\ref{fig3}, our predictions of band gaps with the SHA are in good agreement with the exact calculation of the transfer matrix method. 
Moreover,  dispersion diagrams around the gaps also agree very well. They slowly diverge when moving away from the gaps: the $A$ and $B$ functions are varying faster and thus their second derivatives are not negligible any more. In addition, the development in Eq.~\eqref{eq_u} neglecting harmonics of order higher than 2 and keeping only the main term around wave number $G$ is no longer appropriate. Finally,  the SHA, keeping only one harmonic in Eq.~\eqref{eq_gene}  naturally becomes inadequate when moving too far away from the gap.

Fig.~\ref{fig3} validates our SHA for the first and second band gaps; and it is easy to convince oneself that it is also true for any of the SL band gaps in a first order approximation. Below, we will focus on periodic solutions around the second band gap: SL eigenmodes in the vicinity of this gap will thus be well described by solutions of  Eq.~\eqref{eq_oscill_complx}.

\section{SAM and SLM}
\label{sam-slm}
In the previous section, we have studied the general solution of Eq.~\eqref{eq_gene} and derived the dispersion diagram of the SL\@. 
Since experiments always involve a finite SL,  we now consider the effects of the presence of  a free surface at  $z=0$ with definite boundary conditions on SL eigenmodes. 

\subsection{Semi-infinite medium: surface effects} 
\label{semi_inf}

\begin{figure}
\centerline{\epsfxsize=8cm  \epsfbox{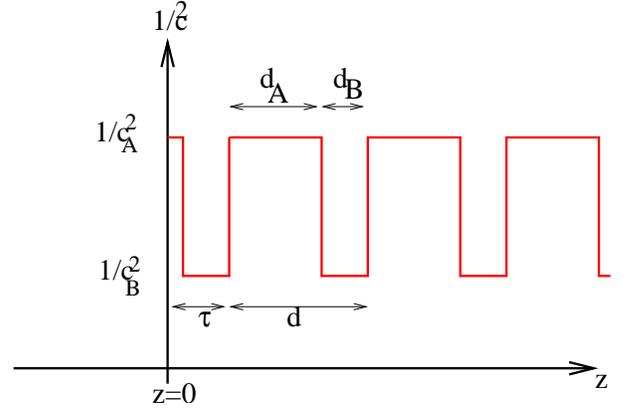} }
\caption{(color online)  Schema of the SL compared to the surface $z=0$}
\label{fig4}
\end{figure}

More precisely,  we intend to study {\it periodic}\/ solutions  of Eq.~\eqref{eq_gene}  in the vicinity of the band gap at $\omega_{BG}^{(2)}$ using  free boundary conditions i.e.\ $u(0)=u_0$ and $\frac{du}{dz}(0)=0$.  In the following, we will use the abbreviation NBPM  (Near Bang gap Propagative Mode) in reference to such modes. Figure~\ref{fig4}  plots $\frac{1}{c^2}$ as a function of z. The dephasing $\tau$ of the SL compared to $z=0$ is defined on Fig.~\ref{fig4}.  As shown below,  $\tau$ will be a key parameter in our study.  
Coefficients $a_0$, $c_m$ and $d_m$ in Eq.~\eqref{eq_gene} are calculated from the Fourier components of  $\frac{1}{c^2}$: 
\begin{subequations}
\begin{align}
&a_0 = 1/\cmoy^2 =\gamma \,  \frac{1}{c^2_A} + \bloc{1-\gamma}\frac{1}{c^2_B}  \label{c0}\\
&c_m =  2 \frac {\frac{\cmoy^2}{{c^2_A}}- \frac{\cmoy^2}{{c^2_B}} } {\pi m   } \cos (\pi m (2 \alpha + \gamma) ) \sin(  \pi m \gamma)   \label{cn}\\
&d_m =  2 \frac {\frac{\cmoy^2}{{c^2_A}}- \frac{\cmoy^2}{{c^2_B}} } {\pi m  }  \sin(  \pi m (2 \alpha + \gamma))  \sin ( \pi m \gamma)  \label{dn} \\
\text{or} \nonumber \\
&p_m =\frac{c_m - i d_m} {2}=  \bloc{\frac{\cmoy^2}{c^2_A}- \frac{\cmoy^2}{c^2_B}} \frac{{e^{-i 2 \pi m \gamma} - 1}}{-i 2 \pi m} e^{-i 2 \pi m \alpha}  \label{pm}
\end{align}
\end{subequations} 
Where $\alpha = \tau/ d$ is the reduced dephasing and $\gamma=d_A/d$ is the cycle ratio of the SL\@.   The mean sound speed $\cmoy$ is defined by Eq.~\eqref{c0}. 

\begin{figure}[h]
\centerline{\epsfxsize=6cm  \epsfbox{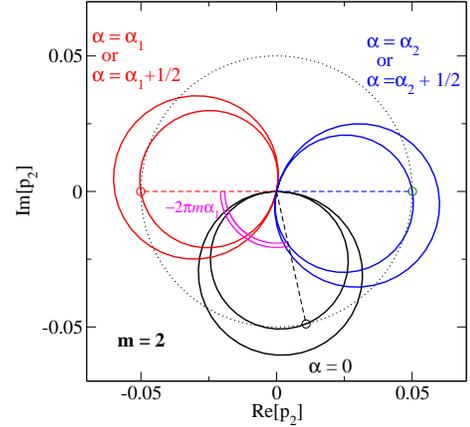} }
\caption{(color online) 
 Plot in the complex plane of $p_2$ \eqref{pm}, for $\gamma \in [0, 1]$. The first case considered is $\alpha = 0$ (black). The curve is not a single circle, since the average sound speed $c_0$ depends on $\gamma$, as seen in Eq.\eqref{c0}.   The reduced dephasing $\alpha$ of the SL merely amounts to a rotation of the curve around 0. The dot circles correspond to the SL taken as an example in this paper with a cycle ratio $\gamma = \gamma_{SL}$.
 For  $\alpha = \alpha_1 = 1/2 - \gamma_{SL}/2$ or $\alpha = \alpha_1 +1/2= 1 - \gamma_{SL}/2$ (red) or $\alpha = \alpha_2 = 3/4 - \gamma_{SL} /2$ or $\alpha = \alpha_2 +1/2 = 5/4 - \gamma_{SL} /2$ (blue), this brings $p_2(\gamma_{SL})$ on the real axis.}
\label{fig5}
\end{figure}

Since we study NBPM in the vicinity of the second band gap around $\omega_{BG}^{(2)}$, it is instructive to consider the evolution of $p_2$ as a function of the cycle ratio $\gamma$  for different values of the reduced dephasing $\alpha$ as shown in Fig.\ref{fig5}. 
$p_2$ vanishes twice as the cycle ratio $\gamma$ varies from 0 to 1. So there are two values of $\gamma$, independent of $\alpha$, for which the SHA predicts no gap opening around $\omega_{BG}^{(2)}$. Note that a gap might actually be opened through the second parametric resonance $n=2$ produced by the first harmonic $m=1$ (cf. Fig.~\ref{fig2}). 

For a given $\gamma$, the effect of $\alpha$ is merely to rotate the figure by $-2\pi m \alpha$, as seen in Eq.\eqref{pm}.
Note that experimentally, most of the SL have $\alpha=0$ and that for $\alpha=0$, $p_2$ as a function of $\gamma$ never cross the real axis except when $p_2=0$: we will illustrate below to what extend the case $p_2$ non null and real is relevant.   

Details of the resolution of Eq.~\eqref{eq_oscill_complx} are given in Appendix~\ref{appen1}.  Since we are focussing on periodic modes, i.e.\ outside the gap,  $\kappa^2$ in Eq.~\eqref{ka2} is positive. Let us introduce $\psi$ the argument of $p_2$: 
\begin{equation}
p_2 = \abs{p_2} \ e^{i \psi}  \label{p2}
\end{equation}
The solution $u(z)$ satisfying the free boundary conditions $u(0)=u_0$ and $\frac{du}{dz}(0)=0$ reads
\begin{eqnarray} 
u(z) = 2 A_0  \left[ ( 1- \zeta)\cos ( G z + \psi/2 ) \cos ( \kappa z + \phi)  \right. \nonumber \\
\left. - (1+ \zeta) \sin ( G z + \psi/2 ) \sin ( \kappa z + \phi) \right] \label{ufin}
\end{eqnarray} 
where $\zeta$, $\phi$  and $A_0$ are defined by
\begin{eqnarray}
\zeta &=& \frac {\delta}{\abs{p_2}} - \sqrt{ \bloc{\frac{\delta}{\abs{p_2}}}^2 - 1} \label{zeta}\\
\tan \phi &=&  - \tan (\psi/2) \frac{  G (1 - \zeta) +  \kappa (1 + \zeta) } {  G (1 + \zeta)  + \kappa (1 - \zeta) } \label{phi} \\
A_0 &=& \frac{u_0}{2\blocc{(1-\zeta) \cos\tfrac{\psi}{2} \cos\phi - (1+\zeta) \sin\tfrac{\psi}{2} \sin\phi}}\label{c}
\end{eqnarray}
$u(z)$ is real and $u(z,t)$ is a stationary wave, as expected since the surface is a perfect mirror for acoustic waves.

Eq.~\eqref{ufin} is linear combination of two terms each involving a quickly oscillating function at $G$ and a slowly one at $\kappa$ corresponding respectively to  Bloch and envelope wave functions.  The parameter $\zeta$ governs the relative contributions of each term  and hence their dominant character which we will discuss in Sec.~\ref{sect_contrast}. The parameter $\phi$ is a dephasing of the envelope function with respect to the SL surface: it will be discussed in Sect.~\ref{sect_phi}. $A_0$ governs the global amplitude of the displacement field.  
There are in general beatings in the solution $u(z)$, as emphasized by the analytical expression for its root mean squared envelope \urms, derived in Appendix~\ref{envelope}:
\begin{equation}
  \urms^2(z) = 2 A_0^2 \blocc{1+\zeta^2 - 2 \zeta \cos2(\kappa z + \phi) }
	\label{urms}
\end{equation}
where the fast oscillating terms at $G$ present in Eq.\eqref{ufin} have been smoothed out.

\subsection{Surface Loving and Avoiding Modes} 
\label{SLAM}
In this section, we discuss the effect of the reduced dephasing $\alpha$ of the SL compared to the surface $z=0$ on NBPM near the second gap around $\omega_{BG}^{(2)}$. 
We especially focus on the phase of the envelope wave function at wave vector $\kappa$. 
Examining the general case requires a numerical study. However, two limiting cases where $p_2$ is real, are remarkable.
\subsubsection{Remarkable cases. Pure SAM and pure SLM}
\label{remarkable}
\begin{itemize}
\item $p_2$ is real and $p_2>0$ ($\alpha=\alpha_2$ or $\alpha_2+1/2$ on Fig.~\ref{fig5}) . Therefore from  \eqref{p2}, $\psi =0$ and from \eqref{phi}, $\phi=0 [\pi]$. 
	\begin{itemize}
	 \item Just above the gap, we have  $\delta \approx \abs{p_2}$ (cf Eqs.~\eqref{ka2} and~\eqref{delta}), 
	implying $\zeta \approx 1$(Eq.~\eqref{zeta}). Thus from Eq.~\eqref{ufin}, the main contribution to $u(z)$ is proportional to $\sin ( G z ) \sin ( \kappa z )$: the envelope of the vibration mode is null at the surface. Following the denomination of Merlin~\cite{merlin_phil_mag}, we call such a mode a  Surface Avoiding Mode (SAM): these SL eigenmodes shy away from the boundaries.  
	\item Just below the gap, $\delta \approx -\abs{p_2}$ and thus $\zeta \approx -1$. The main contribution to $u(z)$ is 
 	$\cos ( G z ) \cos ( \kappa z )$: The envelope amplitude is maximal at the surface. In view of that,  we refer to such SL eigenmode as Surface Loving Mode (SLM). 
	\end{itemize}
\item $p_2$ is real and $p_2<0$ ($\alpha=\alpha_1$ or $\alpha_1+1/2$ on Fig.~\ref{fig5}), $\psi = \pi$ and thus $\phi=  \pi/2 [\pi]$.
	 \begin{itemize}
		 \item Just above the gap, $\delta \approx \abs{p_2}$.  The main contribution to $u(z)$ is then 
			$\sin ( G z + \pi/2) \sin ( \kappa z \pm \pi/2  ) = \pm \cos( G z)\cos(\kappa z)$.  Therefore, the SL eigenmode corresponds to a Surface Loving Mode.
	\item just below the gap   $\delta \approx -\abs{p_2}$. The main contribution to $u(z)$ is 
 			$\cos ( G z + \pi/2 ) \cos ( \kappa z  \pm \pi/2 ) =  \pm \sin ( G z ) \sin ( \kappa z )$: Therefore, the SL eigenmode corresponds to a Surface Avoiding Mode.
	\end{itemize}
\end{itemize} 

Note that we have chosen to base the latter discussion on  Eq.~\eqref{ufin}, but Eq.~\eqref{urms} could also be used and leads to the same conclusions. 

If the distinction between SAM and SLM is obvious in the cases where $p_2$ is real, in the general case we need  a more precise definition:  we will speak about SAM or SLM depending if the envelope amplitude at the surface ($z=0$) is smaller or higher than its value a quarter period later at  $z= \frac{\lambda_\kappa}{4} = \frac{\pi}{2\kappa}$ where  $\lambda_\kappa$ is the wavelength associated with the envelope wave function.  With this definition, according to  Eq.~\eqref{urms}, knowing $\zeta$ and $\phi$ is enough to determine the SAM or SLM character.

\subsubsection{Beatings contrast}
\label{sect_contrast}
\begin{figure}[h]
\centerline{\epsfxsize=6cm  \epsfbox{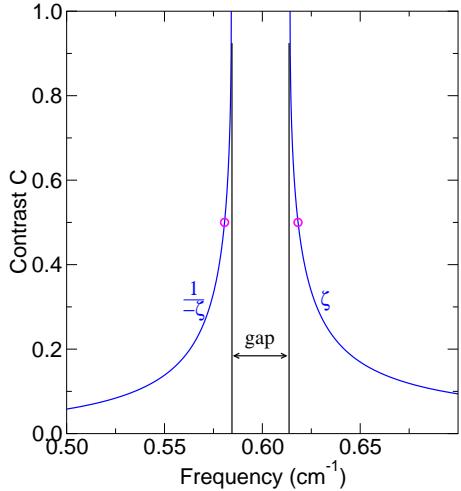} }
\caption{Beating contrast C of the displacement $u$, defined in Eq.\eqref{contrast}  as a function of the frequency $\nu=\frac{\omega}{2 \pi}$. $C$ equals $\zeta$ above the gap and $\frac{1}{-\zeta}$ below. \sayparam \ Dot circles mark a beating contrast of 0.5 for frequencies \nuhalfbelow~THz and \nuhalfabove~THz. }
\label{fig6}
\end{figure}

Let us focus on $\zeta$ which, as previously quoted, governs the relative amplitude of the two contribution in Eq.~\eqref{ufin} and hence drive the beatings contrast, which is defined in Eq.\eqref{contrast} and plotted in Fig.\ref{fig6} as a function of the frequency $\nu=\frac{\omega}{2 \pi}$.
From Fig.~\ref{fig6}, the contrast reaches its maximum value of 1 only at the precise band gap edges. In such a case, the envelope function exactly vanishes at its minimum:  one could then speak about perfect SAM or SLM.  Moving away from the gap,  the contrast tends to 0 giving the same weight to both terms in Eq.~\eqref{ufin}. 
These beatings are also present in an infinite SL, just like in a temporal parametric oscillator. Away from the resonance condition, the excitation goes successively in-phase and out-of-phase with the movement of the system, leading to an increase or a decrease in the amplitude of the oscillations.

\subsubsection{SAM or SLM?}
\label{sect_phi}
To get a strong SAM or SLM, a good beating contrast is needed. Then if the envelope has a minimum or a maximum at the surface, it is a SAM or a SLM, respectively.
Indeed, from Eq.\eqref{urms} the phase of the envelope relative to the SL surface depends essentially on $\phi$ and on the sign of $\zeta$.
The latter is seen in Fig.\ref{fig6} to be negative below the gap and positive above.  In this section we focus on $\phi$, the dephasing of the envelope function.

For a given type of SL (materials and thicknesses), $\phi$ depends essentially on the frequency $\nu=\frac{\omega}{2 \pi}$ and on the reduced dephasing $\alpha$ of  the SL.
As in the whole article, we use here the numerical parameters of the SL described in Sect.~\ref{diag_disp}  whose dispersion diagram is plotted in Fig.~\ref{fig3}. 

\begin{figure}[h]
\centerline{\epsfxsize=8cm  \epsfbox{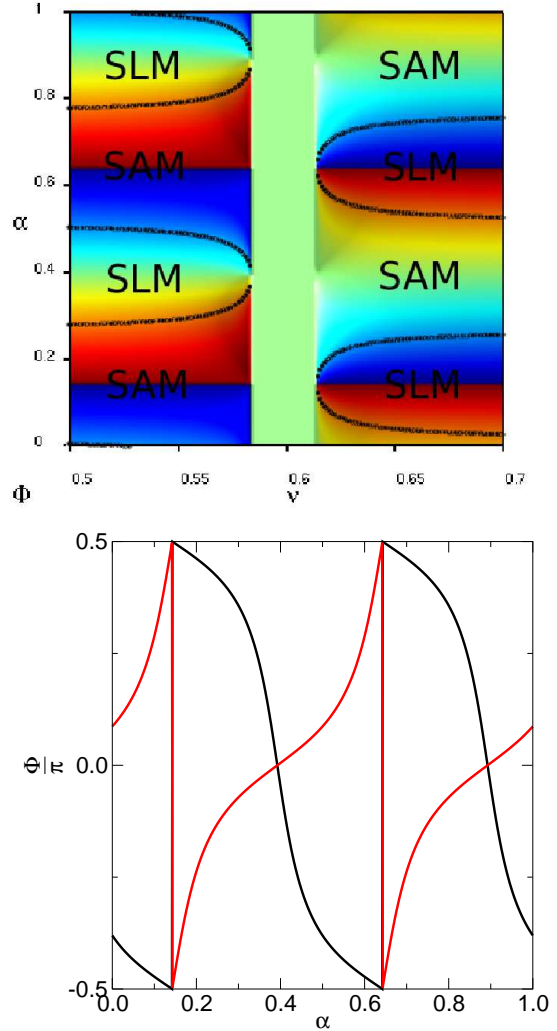} }
\centerline{\epsfxsize=7cm  \epsfbox{fig7b.eps} }
\caption{(color online)  a) Value of $\phi$ (color blue $\phi=-\pi/2$, red $\phi=\pi/2$) as a function of the frequency $\nu$ (THz)  and the reduced dephasing $\alpha$. In the gap, $\phi$ is not defined.\ b) $\phi$  as a function of $\alpha$ at \nuhalfbelow~THz (black) and \nuhalfabove~THz (red). Dark curves separate SAM and SLM regions. These curves have been calculated looking for couples ($\alpha,\nu$) that equal the envelope wave function amplitude in $z=0$ and $z= \frac{\pi}{2\kappa}$, and correspond to $\phi = \pm \frac{\pi}{4}$ (cf Eq.~\eqref{urms}). \sayparam}
\label{fig7}
\end{figure} 
Figure~\ref{fig7}a) presents in a 3D plot the variations of $\phi$ as a function of $\alpha$ and of the frequency $\nu$, calculated using Eq.~\eqref{phi} around the second band gap.  Figure~\ref{fig7}b)  reports  the value of $\phi$ as a function of $\alpha$ for two frequencies above and below the gap: \nuhalfbelow~THz and \nuhalfabove~THz.  These two frequencies reported in Fig.~\ref{fig6}, correspond to a 0.5 beating contrast $C$ (Eq.~\eqref{contrast}),  so that at \nuhalfbelow~THz (resp. \nuhalfabove~THz), the  displacement $u(z)$ (Eq.~\eqref{ufin}) is dominated by the $\cos(G z + \psi/2) \cos(\kappa z + \phi)$ term (resp. $\sin(G z + \psi/2) \sin(\kappa z + \phi)$). 

 From Fig.~\ref{fig7}a) and b),  $\phi$ is a periodic function of $\alpha$  with period $\frac{1}{2}$ at fixed frequency $\nu$: indeed, we are looking at the second band gap  due to the harmonic at wavevector $2 G$; by changing $\alpha$ from 0 to 1, the surface $z=0$ sweeps two periods of that harmonic.
For example,  the two remarkable modes discussed above for which $p_2$ is real, could be obtained for two values of $\alpha$: these values are obtained by cancelling the $d_2$ term in Eq.~\eqref{dn} ($n=2$). This has  also been illustrated in Fig.\ref{fig5}. The case $p_2$ real and positive corresponds to  $\alpha = \alpha_2=  3/4-\gamma/2 \simeq 0.39$ or $\alpha =\alpha_2 +1/2 = 5/4-\gamma/2 \simeq 0.89$,  whereas the case $p_2$ real and negative corresponds to $\alpha = \alpha_1 =1/2-\gamma/2 \simeq 0.14$ or $\alpha = \alpha_1+1/2= 1-\gamma/2 \simeq 0.64$.  

In Fig.~\ref{fig7}a), in addition to the value of $\phi$, we plot  dark curves separating SAM and SLM regions. These curves have been calculated looking for couples ($\alpha,\nu$) that equal the envelope wave function amplitude in $z=0$ and $z= \frac{\pi}{2\kappa}$, and correspond to $\phi = \pm \frac{\pi}{4}$ (cf Eq.~\eqref{urms}).

Below the gap,  Figure~\ref{fig7}a) shows that regions of SAM dominate: especially,  as the frequency  closes with  the gap, the SLM regions are reduced.  This result is also coherent with Fig.~\ref{fig7}b): for \nuhalfbelow~THz, the slope of $\phi$ as a function of $\alpha$ in the vicinity of $\alpha = 3/4-\gamma/2 \simeq 0.39$ is very steep, and would be steeper for a frequency closer to the gap. Above the gap, the same conclusions apply: SAM regions dominate the phase space. 

\subsubsection{SAM/SLM character}
In order to summarize the results of the two previous sections, let us define a ``SAM/SLM character'' as:
\begin{equation}
\chi = \ln \! \left[ \urms(0)/\urms \left(\frac{\pi}{2\kappa}\right) \right]
\label{chi} 
\end{equation}
which is positive for a SLM and negative for a SAM\@. The highest $\abs{\chi}$, the more pronounced is the SAM or SLM character of the mode.
\begin{figure}[h]
\centerline{\epsfxsize=8cm  \epsfbox{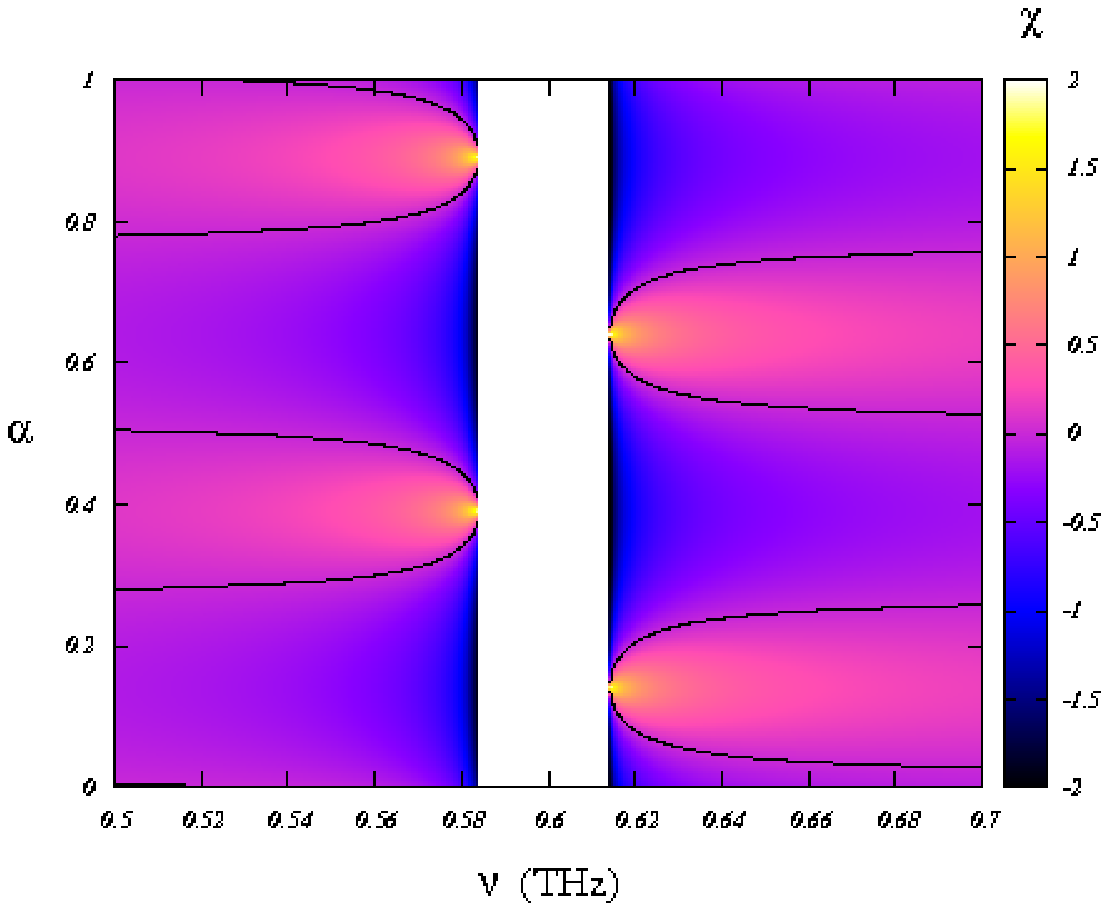} }
\caption{(color online)  SAM/SLM character $\chi$ as defined in Eq.\eqref{chi}. A black curve delimits regions of positive and negative $\chi$. \sayparam}
\label{fig8}
\end{figure}
$\chi$ is plotted as a function of the frequency $\nu$ and the reduced dephasing $\alpha$ of the SL in Fig.\ref{fig8}. It is clear that the more pronounced SAM or SLM (highest $\abs{\chi}$) are localized close to the gap, and that SAM dominate the phase space.

\subsubsection{Numerical Study}
\label{sub_sub_num}

\begin{figure}

\centerline{ \epsfxsize=5.5cm  \epsfbox{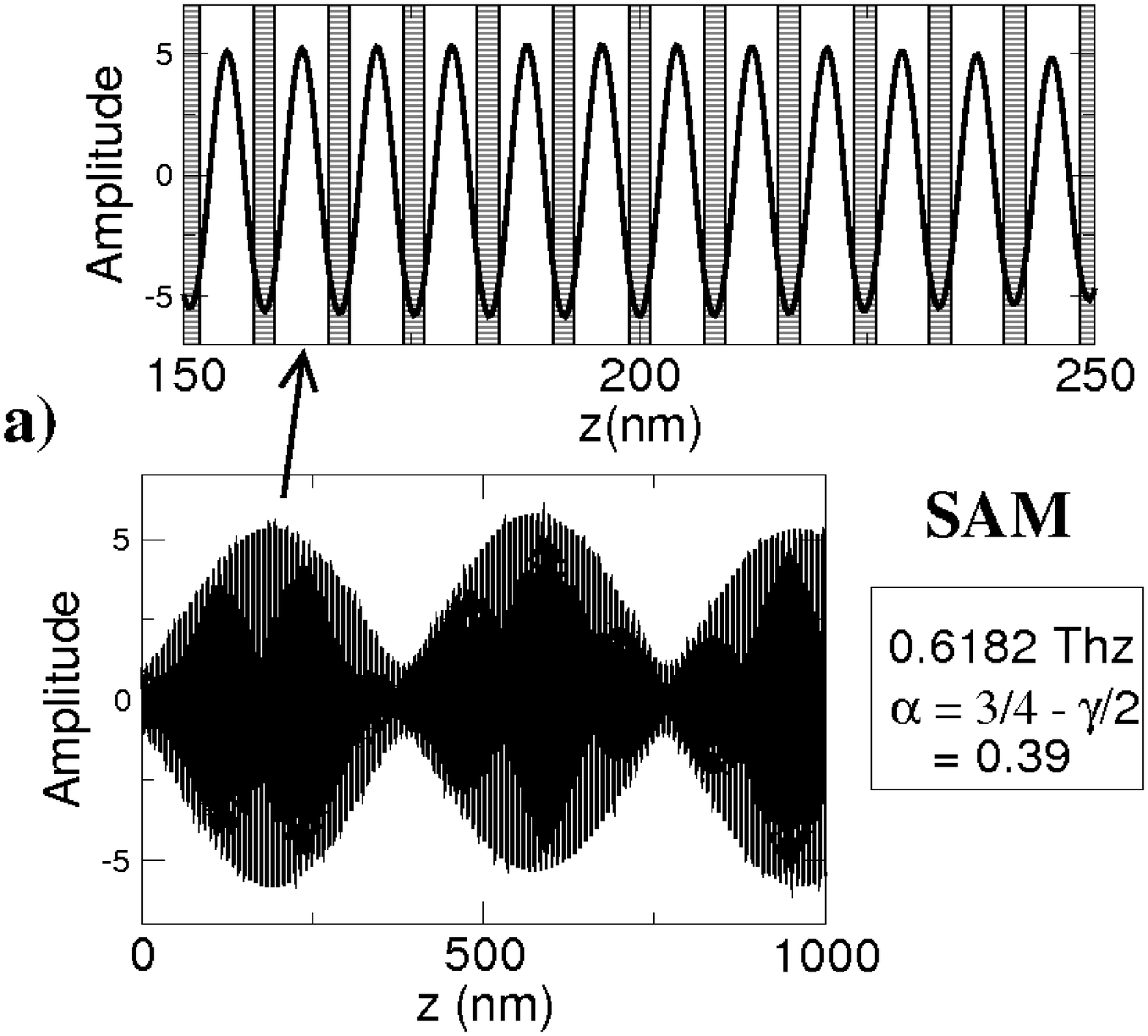} }
\centerline{ \epsfxsize=5.5cm  \epsfbox{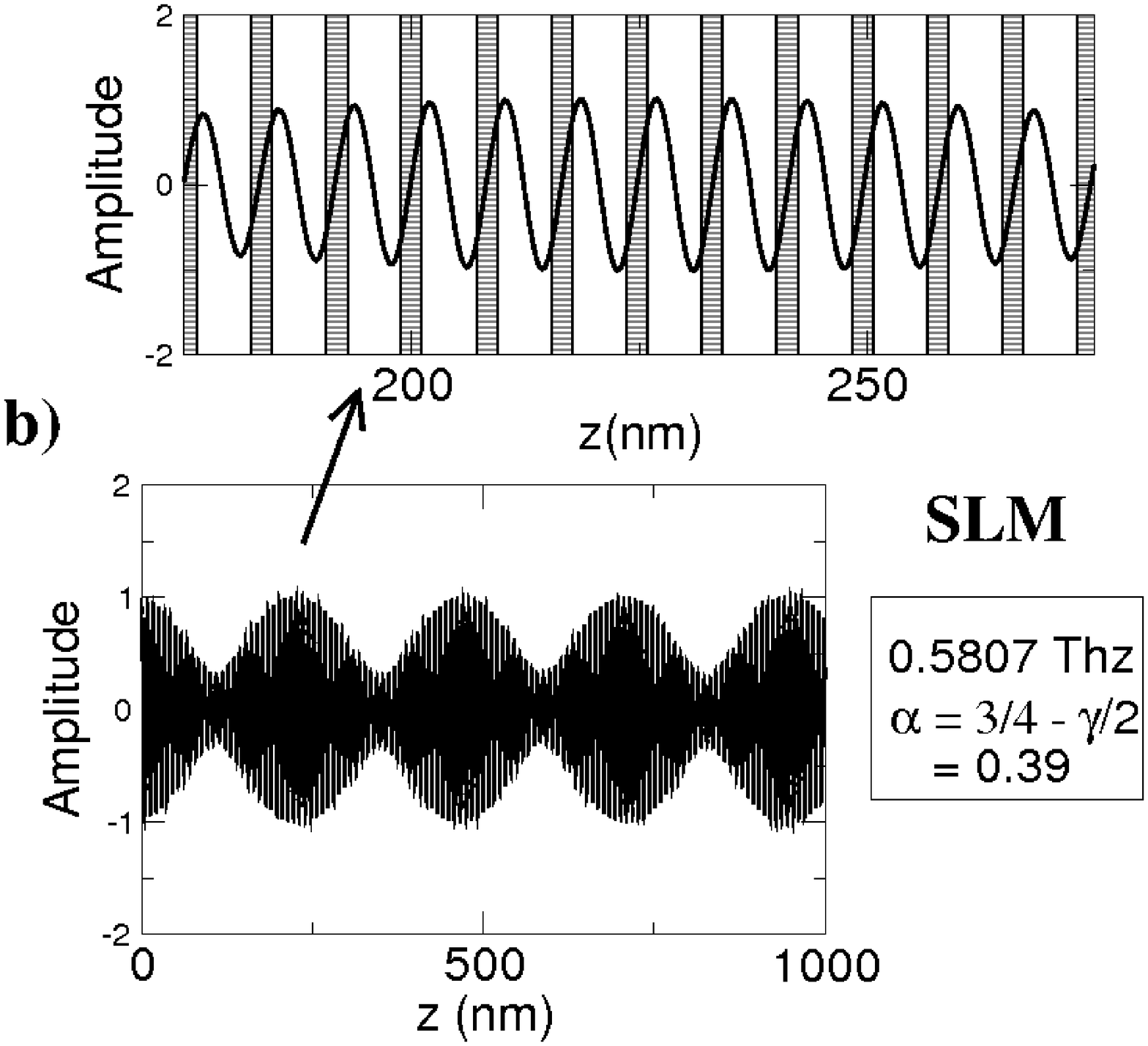} }
\centerline{ \epsfxsize=5.5cm  \epsfbox{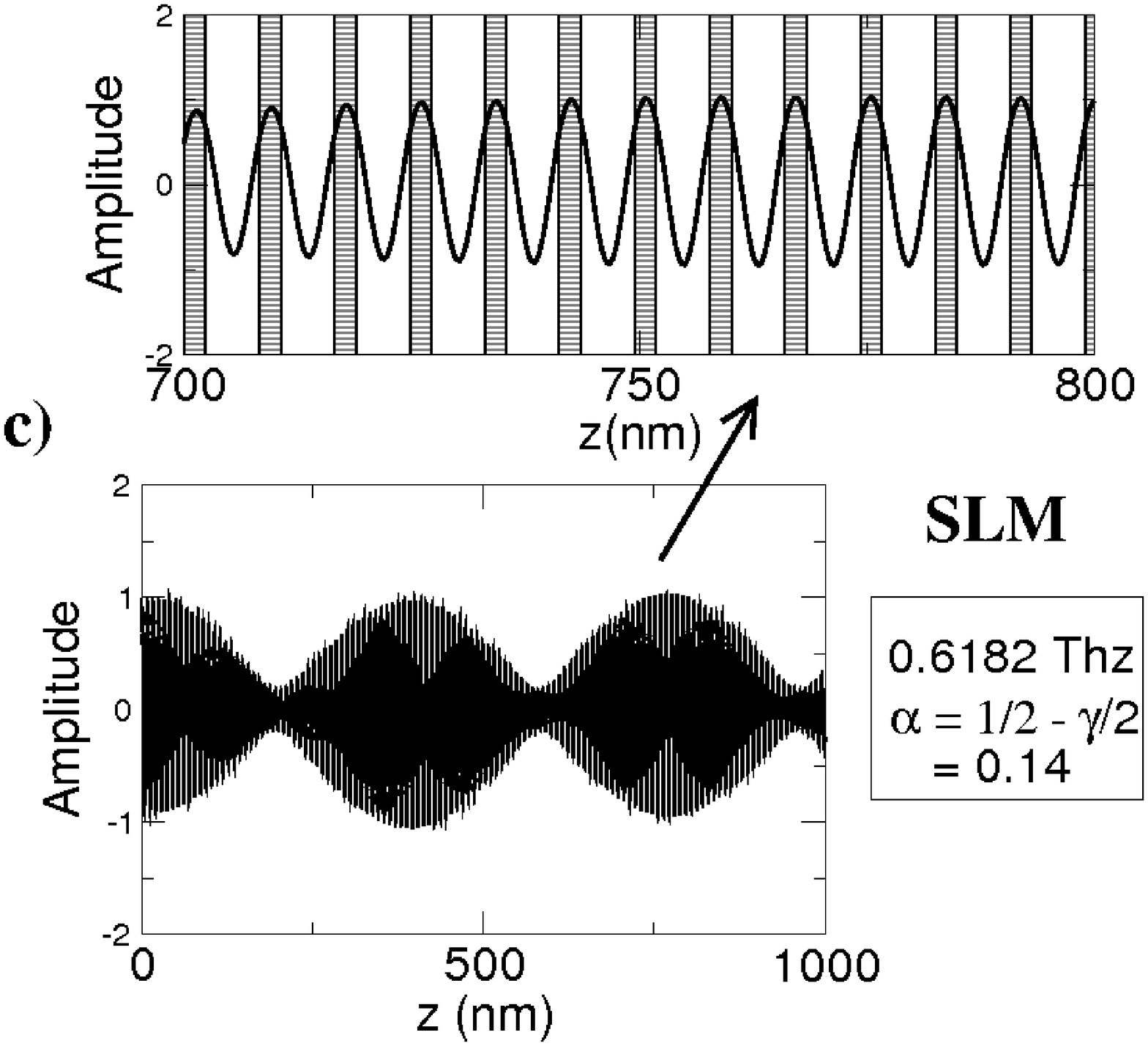} }
\centerline{ \epsfxsize=5.5cm  \epsfbox{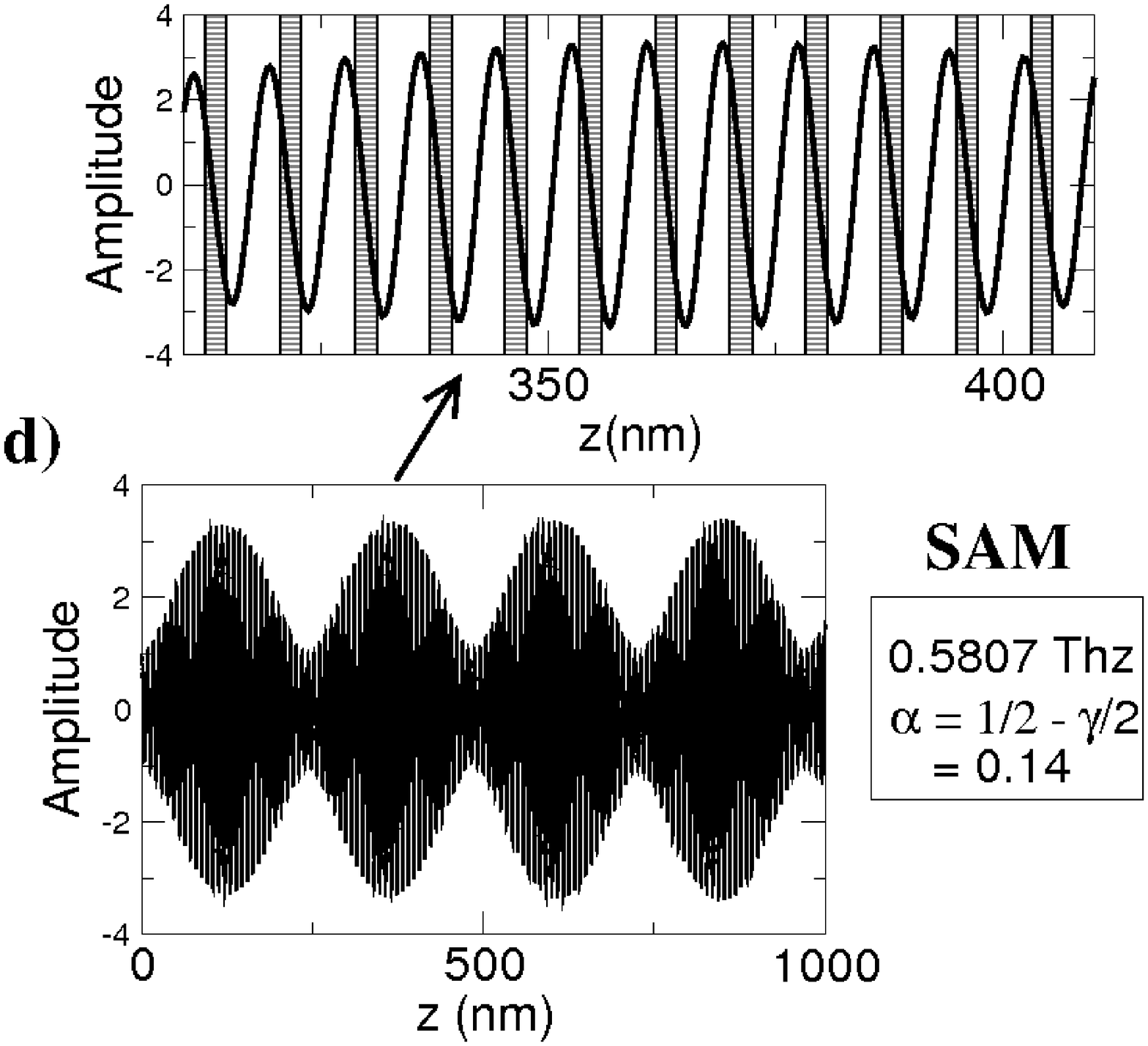} }
\caption{Numerical solutions of Eq.~\eqref{eq_gene} for \nuhalfbelow~THz (below the gap) and \nuhalfabove~THz  (above the gap) with free boundary conditions $u(0)=1$ and $\frac{du}{dz}(0)=0$. The reduced dephasing $\alpha$ of the SL compared to the surface $z=0$ is respectively $\alpha= \alpha_2 = 3/4-\gamma/2 \simeq 0.39$  for figure a) and b) and $\alpha = \alpha_1 = 1/2-\gamma/2\simeq 0.14$ for figure c) and d).  For each graph, we plot both a zoom and overall picture of the mode. The overall picture exhibits the Loving or Avoiding character of the modes. The zoom allows the discussion on the phases of the mode compared to the SL when the amplitude of the mode is maximum. Dashed parts represent AlAs layers and blank ones GaAs layers.  }
\label{fig9}
\end{figure}

To check our analytical predictions based on the SHA, we numerically integrate the differential equation Eq.~\eqref{eq_gene}  using a 7th-order Runge-Kutta integrator around the 2nd band gap: we thus  take into account all harmonics of the inverse square  sound speed.    
We choose here to represent two different cases closed to the remarkable modes we point out above:  $p_2$ real positive for $\alpha = \alpha_2 = 3/4-\gamma/2 \approx 0.39 $ and $p_2$ real negative for $\alpha= \alpha_1 = 1/2-\gamma/2 \approx 0.14$. 
In both cases, we compute solutions of Eq.~\eqref{eq_gene} for \nuhalfbelow~THz and \nuhalfabove~THz (0.5 beating contrast)  with free boundary conditions at the surface, i.e.\ $u(0)=1$ and $\frac{du}{dz}(0)=0$; these numerical results are plotted in Fig.~\ref{fig9}.
For $p_2$ real positive ($\alpha = \alpha_2$), we clearly observe a SAM above the gap and a SLM below, whereas for $p_2$ real negative ($\alpha= \alpha_1$), it is the contrary. With respect to the SAM or SLM character,  the numerical results thus fully agree with our analytical analysis. 

Fig.~\ref{fig9} not only  validates  our calculations and approximations, it also confirms that the parametric resonance analogy is fruitful to get the main physics of wave propagation near band gaps in semi-infinite SL\@. However, please note that numerical values of the contrasts $C$  and the envelope wavelengths measured in Fig.~\ref{fig9} differ from the ones predicted by  Eq.~\eqref{contrast} and  Eq.~\eqref{ka2}.  We attribute these discrepancies to the SHA. Indeed, the SHA allows the prediction of  the width of the 2nd band gap with an error of the order of  $\abs{p_1}^2$. Thus, the gap edges predicted by the parametric oscillator analogy slightly differ from the one calculated from the exact transfer matrix method (or by a direct integration of Eq.~\eqref{eq_gene}). Due to the almost horizontal dispersion diagram curve near the gap edges, a small variation of frequency $\omega$ may induce a non negligible variation in $\kappa$. Consequently, while the parametric oscillator analogy gets the main physics and especially allows us to find out the parameters to see SLM, it remains based on the SHA and thus, some care should be taken in order to predict quantitative values, which will better be derived from an exact approach like the transfer matrix method. 

\section{Observation of SLM and SAM}
\label{observation}

We now address the problem of the experimental observability of NBPM\@. Trigo et al.~\cite{trigo} claim they have shown the existence  of SAM using pump-probe experiments coupled to a theoretical analysis.  In the following, we would like to focus in detail on the optical activity of NBPM.

Colvard et al.~\cite{colvard} addressed the phonon-electron coupling in GaAs/AlAs SL\@. The reduced dephasing  of phonons compared to layers of the SL is a key parameter for the strength of this coupling. They showed in agreement with experiments that, out of electronic resonance conditions,  $B_2$ phonons,  in phase with layers of the SL, do not create a strong electron-phonon coupling and are  thus not observable. On the contrary, $A_1$ phonons, in quadrature phase with layers of  the SL do create a strong electron-phonon coupling and are thus observable.  These conclusions rely on the electron-phonon coupling mechanism: the deformation potential  (in GaAs/AlAs SL), whose hamiltonian is proportional to  the divergence of vibration modes.

In the case of NBPM around the band gap at $\omega_{BG}^{(2)}$, their wave vector $G \pm \kappa$ is slightly different from the primitive vector $G$ of the reciprocal lattice, thus their relative phase compared to layers of the SL slowly shifts with $z$ as $\pm \kappa z$.  Thus, such modes could be in phase or out of phase with the SL depending on the position $z$ in the SL (as long as $\kappa \ne 0$).  However, though the envelope amplitude is varying, we can assume that \emph{these modes will be observable if they have the $A_1$ symmetry (i.e.\ in quadrature phase with the layers of the SL) at their maximum envelope amplitude}. 
  
As demonstrated in Appendix~\ref{appen3} for the remarkable cases of Sect.~\ref{remarkable}, we find that, out of resonance conditions, observable modes are found below the gap independently on their SAM or SLM nature. 

To check our latter analytical  prediction, numerical solutions of Eq.~\eqref{eq_gene} below (\nuhalfbelow~THz) and above (\nuhalfabove~THz)  the gap obtained in Sect.~\ref{sub_sub_num}   are compared to the SL in regions of maximum envelope amplitude.  
Fig.~\ref{fig9} reports these results. In both cases $p_2$ real positive  ($\alpha = 3/4-\gamma/2$) and $p_2$ real negative ($\alpha= 1/2-\gamma/2$), phonons above the gap are in phase with the layers of the SL at their maximum amplitude envelope in agreement with our prediction:  they are hardly observable. On the contrary, phonons below the gap are in quadrature with the layers of the SL and will thus induce a strong electron-phonon coupling in agreement with our predictions.

\section{Discussion} 
\label{discussion}

We have shown that the envelope amplitude of near Brillouin zone center acoustical phonon in the vicinity of the surface can be maximum (SLM) or minimum (SAM) depending on the reduced dephasing of the SL relative to the surface. For a given function $\frac{1}{c^2}(z)$, Eq.~\eqref{eq1} is a second order differential equation whose solutions are entirely determined by two boundary conditions we can reduce to the definition of ($u(0)$,$\frac{du}{dz}(0)$).
Knowing one solution $u_1(z)$,  $u_2(z)=u_1(z+z_{ref})$ is also a solution for the boundary conditions $u_2(0)=u_1(z_{ref})$ and $\frac{du_2}{dz}(0) = \frac{du_1}{dz}(z_{ref})$, and for the function $\frac{1}{c^2}(z+z_{ref})$. Consequently, if SAM exists, a judicious choice of $z_{ref}$ allows to obtain SLM: one just needs to cut the SL at $z_{ref}$ chosen in the region where the envelope of the SAM is maximum. Assuming free boundary conditions, this choice imposes the reduced dephasing $\alpha$ of the SL, which precisely corresponds to our study. Reciprocally, choosing $\alpha=0$, the choice imposes the boundary conditions. 

We now address two discrepancies between our results and related works in the literature.\\
 The first one concerns conclusions of Ref.~\onlinecite{merlin_phil_mag}: ``In the physically important cases of a free and a clamped interface, we emphasize, surface repulsion is a property of all modes with wavevectors close to those of dispersion gaps",  suggesting that only SAM exist. We have shown in this work, that in the case of free boundary conditions, both SAM and SLM exist. The discrepancy between our results and the cited work may be explained  since calculations of Ref.~\onlinecite{merlin_phil_mag} only consider a SL with a reduced dephasing $\alpha=0$ relative to the SL surface. In the precise case of free boundary conditions and $\alpha=0$, our results agree with conclusions of  Ref.~\onlinecite{merlin_phil_mag}.\\
The second discrepancy concerns claims of Ref.~\onlinecite{trigo}: NBPM ``have a tendency to avoid the boundaries, irrespective of the boundary conditions" suggesting that, \emph{whatever} the boundary conditions,  only SAM exist (note that this work~\cite{trigo} also only considers  SL with $\alpha=0$).  If we have restricted our study to the case of free boundary conditions,  of course, the study could easily be extended to the case of any boundary conditions.  Actually, as suggested in the begining of the section, we found that fixing the reduced dephasing $\alpha=0$ and varying the boundary conditions produce the same qualitative conclusions as fixing the boundary conditions and varying the relative phase:  both SAM and SLM exists and  SAM regions dominate the phase space.
 We have chosen not to discuss in detail such effects of changing boundary conditions because the case of free boundary conditions is experimentally the most relevant. To illustrate the correctness of our analysis, and in disagreement with statements of Ref.~\onlinecite{trigo}, the reader can easily check the existence of a SLM by direct integration of Eq.~\eqref{eq_gene} using the following parameters: $u(0)=1.5$ and $\frac{du}{dz}(0)=1$ and $\nu=0.5807$ THz and  the SL defined in Sect.~\ref{diag_disp} with $\alpha=0$ (the same as Ref.~\onlinecite{trigo}).

Let's now consider experimental observations of SAM or SLM\@. 
Ref.~\onlinecite{trigo} claims to  observe SAM under the gap using a SL with $\alpha=0$ and free boundary conditions. This observation agrees with our analysis: from Fig.~\ref{fig8},  NBPM under the gap for $\alpha=0$ are SAM and from Sect.~\ref{observation} they are likely to be observable.  
 
In Sect.~\ref{observation}, we show that both SAM and SLM could in principle be observable. We would like to address here  technical problems associated with the observation of SLM\@. Indeed, in Fig.~\ref{fig8}, we show that SAM regions dominate the phase space. Thus, a high precision on the value of $\alpha$ is needed if one wants to observe SLM\@.  This might prove challenging for SL growth. As an illustration, let's consider the  precision on the thickness of  layers produced by molecular beam epitaxy, which is  about  2 monolayers ($\approx$0.5 nm).  In the case of the SL already studied (AlAs(2.35 nm)/GaAs(5.9 nm))  a precision of $\pm 0.25$~nm on the value of the dephasing $\tau$ corresponds to a precision of $\pm 0.03$ on the value of $\alpha$. 
If $\alpha$ exactly matches $3/4-\gamma/2=0.39$, the NBPM is a SLM for any value of $\nu$ below the gap. However, if  $\alpha$ lays in the range $0.39 \pm 0.03$, 
The frequency $\nu$ has to be lower than  $0.583$~THz to get SLM.  
 Thus, because of the lack of precision on the layers thicknesses, observations of SLM would require a protocol that selects NBPM with a low enough frequency (below the boundary SAM/SLM shown in Fig.~\ref{fig8}): we think such a selection may be achieved by the presence of a substrate. Indeed, the length of the SL and the presence of a substrate is an important parameter that tunes the lifetime of observable phonons created in the SL by the excitation laser. A precise study of these effects will be detailed  in a forthcoming publication.

\section{Conclusion}

We have shown using a fruitful analogy with the parametric oscillator that NBPM could either avoid (SAM) or love the surface (SLM) depending on the relative phase of the SL compared to the sample surface. Moreover, we have shown that both modes should in principle be observable using judicious experimental parameters.  Whereas SAM have already been observed,  experimental evidences of SLM  will be difficult in view of the technical challenge to achieve a convenient SL\@.   
Despite these difficulties, we think such experiments are possible choosing an appropriate SL length. The experimental distinction between SAM and SLM may be demonstrated  using  pump-probe experiments with different penetrating length  lasers. 

  Finally, we would like to underline that our study of sound waves in SL could be generalised to any kind of waves. So that electronic~\cite{ Kucharczyk1992,Steslicka2002}  or electromagnetic~\cite{Scalora1996} wave functions in electronic SL or Bragg mirrors also show a Surface Avoiding or Surface Loving character depending on the surface termination of the SL.

\appendix
 
\section{Resolution of Eq.(6) in a semi-infinite SL} 
 \label{appen1}
Finding the solutions \eqref{eqA} and \eqref{eqB} amounts to diagonalize the square matrix in Eq.~\eqref{mat_AB}. The eigenvalues are $\pm i \kappa$, defined by \eqref{ka2}. Since we are focusing to the outside of the gap, $\kappa$ is real and we may choose it positive (Changing $\kappa$ by $-\kappa$ is equivalent to permute constant $A(B)_+$ and $A(B)_-$ in Eq.~\eqref{eqAB}):
\begin{equation} 
 \kappa = \frac{k_0^2}{2 G} \sqrt{\delta^2 - \abs{p_2}^2 },
\end{equation}
recalling $\delta = \frac{k_0^2 -G^2}{k_0^2}$ from Eq.~\eqref{delta}. $\abs{p_2}$ denotes the magnitude of $p_2$:  $\abs{p_2}=\sqrt{p_2 \bar{p_2}}$.
After diagonalization, the following relations are found between the constants in Eq.~\eqref{eqAB}:
\begin{subequations}
 \begin{align}
  B_+ &= - \zeta e^{-i \psi} A_+\\
  A_- &= - \zeta e^{+i \psi} B_-\\
 \end{align}
\end{subequations}
where $\psi$ is the argument of $p_2$, so that $p_2 = \abs{p_2} e^{i \psi}$ and 
\begin{align}
\zeta &= \frac {\delta}{\abs{p_2}} - \sqrt{\bloc{\frac{\delta}{\abs{p_2}}}^2 - 1}
\end{align}

Let us now impose a real amplitude $u_0$ at the surface, $u(0) = u_0$. This boundary condition reads 
\begin{equation}
 A_+ \bloc{1 - \zeta e^{-i \psi}} + B_- \bloc{1 - \zeta e^{+i \psi}} = u_0
\end{equation}
We may try $B_- = \conj{A_+}$, since their respective factors are complex conjugates, and $u_0$ is real.
Please note that changing $u_0$ to be complex would merely amount to dephase the vibration in time.
Then, introducing $A_0$ (real) and $\phi$ so that $A_+ = A_0 e^{i (\tfrac{\psi}{2}+\phi)}$, the displacement reads
\begin{align}
\label{Aufin}
 u(z) = A_0 &\left \{ \cos\blocc{(G+\kappa)z+\tfrac{\psi}{2}+\phi}- \right.    \nonumber\\
&  \left. \zeta \cos\blocc{(G-\kappa)z+\tfrac{\psi}{2}-\phi} \right\}
\end{align}
with
\begin{equation}
\label{A0}
 A_0 = \frac{u_0}{2\bloc{\cos\bloc{\tfrac{\psi}{2}+\phi} - \zeta \cos\bloc{\tfrac{\psi}{2}-\phi}}}
\end{equation}

The boundary condition $\frac{du}{dz}(0)=0$  yields to the following expression which allows  the determination of the integration constant $\phi$: 
\begin{equation}
\tan\ \phi =  - \tan (\psi/2) \frac{(G+\kappa) - \zeta (G - \kappa) } {(G+\kappa) + \zeta  (G - \kappa) } \label{Aphi}
\end{equation}
Note that Eq.~\eqref{Aphi} defines $\phi$ modulo $\pi$. However,  $u(z)$  is fully defined in Eq.~\eqref{ufin} by defining the couple $A_0$ and $\phi$.
We thus use the convention to choose $A_0>0$ in  Eq.~\eqref{A0}:  $\phi$ is now defined modulo $2 \pi$. Note that close to the gap, this procedure always yields a value of $\phi$ in $]-\frac{\pi}{2}, \frac{\pi}{2}]$.

We found one solution \eqref{Aufin} of this second order differential equation. This is the only one satisfying the given free boundary conditions. 
Finally, Eq.~\eqref{Aufin} can be  written in the alternative form \eqref{ufin}, more convenient for the interpretation.

\section{Envelope equation} 
\label{envelope}
As seen in Fig.\ref{fig9}, the displacement consist in a fast oscillating term at wave vector $G$, modulated by a lower wave vector envelope. One way to define this envelope is to take the square of $u$ in Eq.\eqref{ufin} 
\begin{align}
 \begin{split}
  u^2(z) = 4 A_0^2 \left\{(1-\zeta)^2 \cos^2(Gz+\tfrac{\psi}{2}) \cos^2(\kappa z + \phi) + \right.   \\
  (1+\zeta)^2 \sin^2(Gz+\tfrac{\psi}{2}) \sin^2(\kappa z + \phi) - \\
  (1-\zeta) (1+\zeta) 2  \cos(Gz+\tfrac{\psi}{2}) \sin(Gz+\tfrac{\psi}{2})\\
\left.  \cos(\kappa z + \phi)  \sin(\kappa z + \phi) \phantom{\sin^2(\tfrac{\psi}{2})} \right \} \label{u2_rms}
 \end{split}
\end{align}
and then to average on the characteristic time of the fast oscillating term.   So, we define: 
\begin{equation} 
\urms^2(z) = \frac{G}{2 \pi} \int_{z-\frac{2 \pi}{G}}^{z+\frac{2 \pi}{G} } u^2(z') dz'
\end{equation} 
This procedure  amounts to take a local mean, around every position $z$, on a $\frac{2 \pi}{G}$ range.
In Eq.~\eqref{u2_rms},  using $\kappa \ll G$, the $\cos^2(G z+\tfrac{\psi}{2})$ and $ \sin^2(Gz+\tfrac{\psi}{2})$ yield $\frac{1}{2}$ and the last term vanishes. Finally, we end up with a local ``root mean square'' positive amplitude \urms
\begin{equation}
  \urms^2(z) \simeq 2 A_0^2 \blocc{1+\zeta^2 - 2 \zeta \cos2(\kappa z + \phi) }
\end{equation}
which will also be called the envelope of the displacement in this paper.
The contrast of the beatings in $u$ may be defined as
\begin{equation}
 C = \frac{\max(\urms) - \min(\urms)}{\max(\urms) + \min(\urms)}
\label{contrast}
\end{equation}
which is equal to $\zeta$ above the gap and to $\frac{1}{-\zeta}$ below the gap, since $\max(\urms) = \sqrt{2} A_0 (1 + \abs{\zeta})$ and $\min(\urms) = \sqrt{2} A_0 \abs{1- \abs{\zeta}}$, with $\zeta \in ]0, 1]$ above the gap and $\zeta \in ]-\infty, -1]$ below the gap.

\section{Observation of SAM or SLM: Analytical study}
\label{appen3}

To check if NBPM around band gap at $\omega_{BG}^{(2)}$ have the $A_1$ symmetry  at their maximum envelope amplitude,  
we examine their relative phase compared to the layers of the SL. So that, whereas  in Sect.~\ref{SLAM} we were interested in the phase $\phi$ of the envelope compared to the SL surface, we now get interested in the phase $\frac{\psi}{2}$ of the quickly oscillating function (at spatial frequency G) compared to the layers of the SL.  \\

Eq.~\eqref{ufin} gives the analytical expression of NBPM\@. The layers of the SL can be described   using  
the inverse square sound speed as shown in Fig.~\ref{fig4}. However, only the first harmonic (at $G$) of the SL is necessary to get a picture of the layers of the SL: the coefficient  $p_1$ defined  by Eq.~\eqref{pm} determines the phase of layers of the SL.

We thus, only have to compare the relative phase of the phonons (at their maximum envelope amplitude) to the one of the first harmonic of the inverse square sound speed.

We now examine the two remarkable cases mentioned in Sect~\ref{remarkable}.

Let us start with the cases $p_2$ real and  positive obtained for $\alpha = 3/4-\gamma/2$ or $\alpha= 5/4-\gamma/2$: the SAM mode,
above the gap, varies like $\sin(Gz) \sin(\kappa z)$, whereas the SLM, below the gap, varies like $\cos (G z) \cos (\kappa z)$.
From Eq.~\eqref{cn} and Eq.~\eqref{dn}, we can see that $c_1=0$ and $d_1 \neq 0$ which implies that the first harmonic of the SL varies like $\sin(G z)$. 
Thus, it turns out that the SAM mode is in phase with the
SL at its maximum amplitude ($kz = \pi/2[\pi])$): the electron-phonon
coupling is thus weak which makes it hardly observable. On the
contrary the SLM mode is in quadrature with the SL at its maximum
amplitude ($kz =0[\pi])$): the electron-phonon coupling is high and
 so, this SLM mode is likely to be observable.

A similar analysis for the cases $p_2$ real and negative ($\alpha = 1/2-\gamma/2$ or $\alpha= 3/2-\gamma/2$)
leads to the conclusion that in that case, the SLM mode, above the gap,
will be hardly observable whereas the SAM mode just below the gap, is likely to be observable.

Hence, in both remarkable cases, observable modes are found below the gap.


\begin{thebibliography}{27}
\expandafter\ifx\csname natexlab\endcsname\relax\def\natexlab#1{#1}\fi
\expandafter\ifx\csname bibnamefont\endcsname\relax
  \def\bibnamefont#1{#1}\fi
\expandafter\ifx\csname bibfnamefont\endcsname\relax
  \def\bibfnamefont#1{#1}\fi
\expandafter\ifx\csname citenamefont\endcsname\relax
  \def\citenamefont#1{#1}\fi
\expandafter\ifx\csname url\endcsname\relax
  \def\url#1{\texttt{#1}}\fi
\expandafter\ifx\csname urlprefix\endcsname\relax\def\urlprefix{URL }\fi
\providecommand{\bibinfo}[2]{#2}
\providecommand{\eprint}[2][]{\url{#2}}

\bibitem[{\citenamefont{Brillouin}(1946)}]{brillouin}
\bibinfo{author}{\bibfnamefont{L.}~\bibnamefont{Brillouin}},
  \emph{\bibinfo{title}{Wave Propagation in Periodic Structures}}
  (\bibinfo{publisher}{Dover Publications Inc.}, \bibinfo{year}{1946}).

\bibitem[{\citenamefont{Ashcroft and Mermin}(1976)}]{ashcroft}
\bibinfo{author}{\bibfnamefont{N.}~\bibnamefont{Ashcroft}} \bibnamefont{and}
  \bibinfo{author}{\bibfnamefont{N.~D.} \bibnamefont{Mermin}},
  \emph{\bibinfo{title}{Solid State Physics}} (\bibinfo{publisher}{Holt,
  Rinehart and Winston}, \bibinfo{year}{1976}).

\bibitem[{\citenamefont{Yablonovitch}(1987)}]{yablonovitch}
\bibinfo{author}{\bibfnamefont{E.}~\bibnamefont{Yablonovitch}},
  \bibinfo{journal}{Phys. Rev. Lett.} \textbf{\bibinfo{volume}{58}},
  \bibinfo{pages}{2059} (\bibinfo{year}{1987}).

\bibitem[{\citenamefont{John}(1987)}]{john}
\bibinfo{author}{\bibfnamefont{S.}~\bibnamefont{John}}, \bibinfo{journal}{Phys.
  Rev. Lett.} \textbf{\bibinfo{volume}{58}}, \bibinfo{pages}{2486}
  (\bibinfo{year}{1987}).

\bibitem[{\citenamefont{PascualWinter et~al.}(2007)\citenamefont{PascualWinter, Rozas, Fainstein, Jusserand, Perrin, Huynh, Vaccaro, and Saravanan}}]{winter}
\bibinfo{author}{\bibfnamefont{M.~F.}~\bibnamefont{PascualWinter}},
  \bibinfo{author}{\bibfnamefont{G.}~\bibnamefont{Rozas}},
  \bibinfo{author}{\bibfnamefont{A.}~\bibnamefont{Fainstein}},
  \bibinfo{author}{\bibfnamefont{B.}~\bibnamefont{Jusserand}},
  \bibinfo{author}{\bibfnamefont{B.}~\bibnamefont{Perrin}},
  \bibinfo{author}{\bibfnamefont{A.}~\bibnamefont{Huynh}},
  \bibinfo{author}{\bibfnamefont{P.~O.} \bibnamefont{Vaccaro}},
  \bibnamefont{and}
  \bibinfo{author}{\bibfnamefont{S.}~\bibnamefont{Saravanan}},
  \bibinfo{journal}{Phys. Rev. Lett.} \textbf{\bibinfo{volume}{98}},
  \bibinfo{eid}{265501} (pages~\bibinfo{numpages}{4}) (\bibinfo{year}{2007}).

\bibitem[{\citenamefont{Kent et~al.}(2006)\citenamefont{Kent, Kini, Stanton,
  Henini, Glavin, Kochelap, and Linnik}}]{kent2006}
\bibinfo{author}{\bibfnamefont{A.~J.} \bibnamefont{Kent}},
  \bibinfo{author}{\bibfnamefont{R.~N.} \bibnamefont{Kini}},
  \bibinfo{author}{\bibfnamefont{N.~M.} \bibnamefont{Stanton}},
  \bibinfo{author}{\bibfnamefont{M.}~\bibnamefont{Henini}},
  \bibinfo{author}{\bibfnamefont{B.~A.} \bibnamefont{Glavin}},
  \bibinfo{author}{\bibfnamefont{V.~A.} \bibnamefont{Kochelap}},
  \bibnamefont{and} \bibinfo{author}{\bibfnamefont{T.~L.}
  \bibnamefont{Linnik}}, \bibinfo{journal}{Phys. Rev. Lett.}
  \textbf{\bibinfo{volume}{96}}, \bibinfo{eid}{215504}
  (pages~\bibinfo{numpages}{4}) (\bibinfo{year}{2006}).

\bibitem[{\citenamefont{Tamura et~al.}(2005)\citenamefont{ichiro Tamura,
  Watanabe, and Kawasaki}}]{tamura:165306}
\bibinfo{author}{\bibfnamefont{S.~I}~\bibnamefont{Tamura}},
  \bibinfo{author}{\bibfnamefont{H.}~\bibnamefont{Watanabe}}, \bibnamefont{and}
  \bibinfo{author}{\bibfnamefont{T.}~\bibnamefont{Kawasaki}},
  \bibinfo{journal}{Phys. Rev. B}
  \textbf{\bibinfo{volume}{72}}, \bibinfo{eid}{165306}
  (pages~\bibinfo{numpages}{11}) (\bibinfo{year}{2005}).

\bibitem[{\citenamefont{Gorishnyy et~al.}(2005)\citenamefont{Gorishnyy,
  Maldovan, Ullal, and Thomas}}]{gorishny}
\bibinfo{author}{\bibfnamefont{T.}~\bibnamefont{Gorishnyy}},
  \bibinfo{author}{\bibfnamefont{M.}~\bibnamefont{Maldovan}},
  \bibinfo{author}{\bibfnamefont{C.}~\bibnamefont{Ullal}}, \bibnamefont{and}
  \bibinfo{author}{\bibfnamefont{E.}~\bibnamefont{Thomas}},
  \bibinfo{journal}{physics world} \textbf{\bibinfo{volume}{18}},
  \bibinfo{pages}{24} (\bibinfo{year}{2005}).

\bibitem[{\citenamefont{Mart\'inez-Sala
  et~al.}(1995)\citenamefont{Mart\'inez-Sala, Sancho, S\'anchez, G\'omez,
  Llinares, and Meseguer}}]{meseguer}
\bibinfo{author}{\bibfnamefont{R.}~\bibnamefont{Mart\'inez-Sala}},
  \bibinfo{author}{\bibfnamefont{J.}~\bibnamefont{Sancho}},
  \bibinfo{author}{\bibfnamefont{J.~V.} \bibnamefont{S\'anchez}},
  \bibinfo{author}{\bibfnamefont{V.}~\bibnamefont{G\'omez}},
  \bibinfo{author}{\bibfnamefont{J.}~\bibnamefont{Llinares}}, \bibnamefont{and}
  \bibinfo{author}{\bibfnamefont{F.}~\bibnamefont{Meseguer}},
  \bibinfo{journal}{Nature} \textbf{\bibinfo{volume}{378}},
  \bibinfo{pages}{241} (\bibinfo{year}{1995}).

\bibitem[{\citenamefont{Tamm}(1932)}]{tamm}
\bibinfo{author}{\bibfnamefont{I.}~\bibnamefont{Tamm}}, \bibinfo{journal}{Phys.
  Z. Sowjetunion} \textbf{\bibinfo{volume}{1}}, \bibinfo{pages}{733}
  (\bibinfo{year}{1932}).

\bibitem[{\citenamefont{Rytov}(1956)}]{rytov}
\bibinfo{author}{\bibfnamefont{S.~M.} \bibnamefont{Rytov}},
  \bibinfo{journal}{Akust. Zh.} \textbf{\bibinfo{volume}{2}},
  \bibinfo{pages}{71} (\bibinfo{year}{1956}).

\bibitem[{\citenamefont{Colvard et~al.}(1980)\citenamefont{Colvard, Merlin,
  Klein, and Gossard}}]{colvard}
\bibinfo{author}{\bibfnamefont{C.}~\bibnamefont{Colvard}},
  \bibinfo{author}{\bibfnamefont{R.}~\bibnamefont{Merlin}},
  \bibinfo{author}{\bibfnamefont{M.~V.} \bibnamefont{Klein}}, \bibnamefont{and}
  \bibinfo{author}{\bibfnamefont{A.~C.} \bibnamefont{Gossard}},
  \bibinfo{journal}{Phys. Rev. Lett.} \textbf{\bibinfo{volume}{45}},
  \bibinfo{pages}{298} (\bibinfo{year}{1980}).

\bibitem[{\citenamefont{Jusserand et~al.}(1987)\citenamefont{Jusserand, Paquet,
  Mollot, Alexandre, and Le~Roux}}]{jusserand87}
\bibinfo{author}{\bibfnamefont{B.}~\bibnamefont{Jusserand}},
  \bibinfo{author}{\bibfnamefont{D.}~\bibnamefont{Paquet}},
  \bibinfo{author}{\bibfnamefont{F.}~\bibnamefont{Mollot}},
  \bibinfo{author}{\bibfnamefont{F.}~\bibnamefont{Alexandre}},
  \bibnamefont{and} \bibinfo{author}{\bibfnamefont{G.}~\bibnamefont{Le~Roux}},
  \bibinfo{journal}{Phys. Rev. B} \textbf{\bibinfo{volume}{35}},
  \bibinfo{pages}{2808} (\bibinfo{year}{1987}).

\bibitem[{\citenamefont{Bartels et~al.}(1999)\citenamefont{Bartels, Dekorsy,
  Kurz, and K\"ohler}}]{bartels}
\bibinfo{author}{\bibfnamefont{A.}~\bibnamefont{Bartels}},
  \bibinfo{author}{\bibfnamefont{T.}~\bibnamefont{Dekorsy}},
  \bibinfo{author}{\bibfnamefont{H.}~\bibnamefont{Kurz}}, \bibnamefont{and}
  \bibinfo{author}{\bibfnamefont{K.}~\bibnamefont{K\"ohler}},
  \bibinfo{journal}{Physica B: Condensed Matter}
  \textbf{\bibinfo{volume}{263-264}}, \bibinfo{pages}{45}
  (\bibinfo{year}{1999}).

\bibitem[{\citenamefont{Mizoguchi et~al.}(1999)\citenamefont{Mizoguchi, Hase,
  Nakashima, and Nakayama}}]{mizoguchi}
\bibinfo{author}{\bibfnamefont{K.}~\bibnamefont{Mizoguchi}},
  \bibinfo{author}{\bibfnamefont{M.}~\bibnamefont{Hase}},
  \bibinfo{author}{\bibfnamefont{S.}~\bibnamefont{Nakashima}},
  \bibnamefont{and} \bibinfo{author}{\bibfnamefont{M.}~\bibnamefont{Nakayama}},
  \bibinfo{journal}{Phys. Rev. B} \textbf{\bibinfo{volume}{60}},
  \bibinfo{pages}{8262} (\bibinfo{year}{1999}).

\bibitem[{\citenamefont{Takeuchi et~al.}(2002)\citenamefont{Takeuchi,
  Mizoguchi, Hino, and Nakayama}}]{takeuchi}
\bibinfo{author}{\bibfnamefont{H.}~\bibnamefont{Takeuchi}},
  \bibinfo{author}{\bibfnamefont{K.}~\bibnamefont{Mizoguchi}},
  \bibinfo{author}{\bibfnamefont{T.}~\bibnamefont{Hino}}, \bibnamefont{and}
  \bibinfo{author}{\bibfnamefont{M.}~\bibnamefont{Nakayama}},
  \bibinfo{journal}{Physica B: condensed Matter}
  \textbf{\bibinfo{volume}{316-317}}, \bibinfo{pages}{308}
  (\bibinfo{year}{2002}).

\bibitem[{\citenamefont{Hudert et~al.}(2007)\citenamefont{Hudert, Bartels,
  Janke, Dekorsy, and K{\"{o}}hler}}]{hudert}
\bibinfo{author}{\bibfnamefont{F.}~\bibnamefont{Hudert}},
  \bibinfo{author}{\bibfnamefont{A.}~\bibnamefont{Bartels}},
  \bibinfo{author}{\bibfnamefont{C.}~\bibnamefont{Janke}},
  \bibinfo{author}{\bibfnamefont{T.}~\bibnamefont{Dekorsy}}, \bibnamefont{and}
  \bibinfo{author}{\bibfnamefont{K.}~\bibnamefont{K{\"{o}}hler}},
  \bibinfo{journal}{Journal of Physics: Conference Series}
  \textbf{\bibinfo{volume}{92}}, \bibinfo{pages}{012012}
  (\bibinfo{year}{2007}).

\bibitem[{\citenamefont{Trigo et~al.}(2006)\citenamefont{Trigo, Eckhause,
  Reason, Goldman, and Merlin}}]{trigo}
\bibinfo{author}{\bibfnamefont{M.}~\bibnamefont{Trigo}},
  \bibinfo{author}{\bibfnamefont{T.~A.} \bibnamefont{Eckhause}},
  \bibinfo{author}{\bibfnamefont{M.}~\bibnamefont{Reason}},
  \bibinfo{author}{\bibfnamefont{R.~S.} \bibnamefont{Goldman}},
  \bibnamefont{and} \bibinfo{author}{\bibfnamefont{R.}~\bibnamefont{Merlin}},
  \bibinfo{journal}{Phys. Rev. Lett.} \textbf{\bibinfo{volume}{97}},
  \bibinfo{pages}{124301} (\bibinfo{year}{2006}).

\bibitem[{\citenamefont{Scalora et~al.}(2006)\citenamefont{Scalora, Flynn, Reinhardt, Fork, Bloemer,
 Tocci, Bowden, Ledbetter, Bendickson, Dowling and Leavitt}}]{Scalora1996}
\bibinfo{author}{\bibfnamefont{M.}~\bibnamefont{Scalora}},
  \bibinfo{author}{\bibfnamefont{R. ~J.} \bibnamefont{Flynn}},
  \bibinfo{author}{\bibfnamefont{S.~B.}~\bibnamefont{Reinhardt}},
  \bibinfo{author}{\bibfnamefont{R.~L.} \bibnamefont{Fork}},
  \bibinfo{author}{\bibfnamefont{M.~J.} \bibnamefont{Bloemer}},
\bibinfo{author}{\bibfnamefont{M.~D.} \bibnamefont{Tocci}},	
\bibinfo{author}{\bibfnamefont{C.~M.} \bibnamefont{Bowden}},
\bibinfo{author}{\bibfnamefont{H.~S.} \bibnamefont{Ledbetter}},
\bibinfo{author}{\bibfnamefont{J.~M.} \bibnamefont{Bendickson}},
\bibinfo{author}{\bibfnamefont{J.~P.} \bibnamefont{Dowling}},
 \bibnamefont{and} \bibinfo{author}{\bibfnamefont{R.~P.}~\bibnamefont{Leavitt}},
  \bibinfo{journal}{Phys. Rev. E} \textbf{\bibinfo{volume}{54}},
  \bibinfo{pages}{R1078} (\bibinfo{year}{1996}).

\bibitem[{\citenamefont{Aynaou et~al.}(1988)\citenamefont{Aynaou, El Boudouti, Djafari-Rouhani, Akjouj and  Velasco}}]{Aynaou2005}
\bibinfo{author}{\bibfnamefont{H.}~\bibnamefont{Aynaou}},
\bibinfo{author}{\bibfnamefont{E.~H.}~\bibnamefont{El Boudouti}},
  \bibinfo{author}{\bibfnamefont{B.}~\bibnamefont{Djafari-Rouhani}},
  \bibinfo{author}{\bibfnamefont{A.}~\bibnamefont{Akjouj}},
  \bibnamefont{and} \bibinfo{author}{\bibfnamefont{V.~R..}~\bibnamefont{Velasco}},
  \bibinfo{journal}{J. Phys.:Condens. Matter} \textbf{\bibinfo{volume}{17}},
  \bibinfo{pages}{4245} (\bibinfo{year}{2005}).


\bibitem[{\citenamefont{He et~al.}(1988)\citenamefont{He, Djafari-Rouhani, and
  Sapriel}}]{he}
\bibinfo{author}{\bibfnamefont{J.}~\bibnamefont{He}},
  \bibinfo{author}{\bibfnamefont{B.}~\bibnamefont{Djafari-Rouhani}},
  \bibnamefont{and} \bibinfo{author}{\bibfnamefont{J.}~\bibnamefont{Sapriel}},
  \bibinfo{journal}{Phys. Rev. B} \textbf{\bibinfo{volume}{37}},
  \bibinfo{pages}{4086} (\bibinfo{year}{1988}).

\bibitem[{\citenamefont{Kushwaha et~al.}(1993)\citenamefont{Kushwaha, Halevi,
  Dobrzynski, and Djafari-Rouhani}}]{kushwaha}
\bibinfo{author}{\bibfnamefont{M.~S.} \bibnamefont{Kushwaha}},
  \bibinfo{author}{\bibfnamefont{P.}~\bibnamefont{Halevi}},
  \bibinfo{author}{\bibfnamefont{L.}~\bibnamefont{Dobrzynski}},
  \bibnamefont{and}
  \bibinfo{author}{\bibfnamefont{B.}~\bibnamefont{Djafari-Rouhani}},
  \bibinfo{journal}{Phys. Rev. Lett.} \textbf{\bibinfo{volume}{71}},
  \bibinfo{pages}{2022} (\bibinfo{year}{1993}).

\bibitem[{\citenamefont{Combe~al.}(2008)\citenamefont{Combe, Huntzinger, and Morillo}}]{Combe2008}
\bibinfo{author}{\bibfnamefont{N.} \bibnamefont{Combe}},
  \bibinfo{author}{\bibfnamefont{J.~R.}~\bibnamefont{Huntzinger}},
  \bibnamefont{and}
  \bibinfo{author}{\bibfnamefont{J.}~\bibnamefont{Morillo}},
  \bibinfo{journal}{Bulletin de l'Union des Physiciens} \textbf{\bibinfo{volume}{908}},
  \bibinfo{pages}{133} (\bibinfo{year}{2008}).


\bibitem[{\citenamefont{Blanch}(1972)}]{Abramowitz}
\bibinfo{author}{\bibfnamefont{G.}~\bibnamefont{Blanch}},
  \emph{\bibinfo{title}{Handbook of Mathematical Functions}}
  (\bibinfo{year}{1972}), chap.~\bibinfo{chapter}{20}, p. \bibinfo{pages}{722}.

\bibitem[{\citenamefont{Landau and Lifchitz}(1969)}]{landau_meca}
\bibinfo{author}{\bibfnamefont{L.}~\bibnamefont{Landau}} \bibnamefont{and}
  \bibinfo{author}{\bibfnamefont{E.}~\bibnamefont{Lifchitz}},
  \emph{\bibinfo{title}{Mechanics}} (\bibinfo{publisher}{Mir},
  \bibinfo{address}{Moscow}, \bibinfo{year}{1969}), \bibinfo{edition}{3rd} ed.

\bibitem[{\citenamefont{Merlin}(1994)}]{merlin_phil_mag}
\bibinfo{author}{\bibfnamefont{R.}~\bibnamefont{Merlin}},
  \bibinfo{journal}{Philos. Mag. B} \textbf{\bibinfo{volume}{70}},
  \bibinfo{pages}{761} (\bibinfo{year}{1994}).

\bibitem[{\citenamefont{Kucharczyk and M. Steslicka}(1992)}]{Kucharczyk1992}
\bibinfo{author}{\bibfnamefont{R.}~\bibnamefont{Kucharczyk}}, 
\bibnamefont{and}  \bibinfo{author}{\bibfnamefont{M.}~\bibnamefont{St\c{e}\'slicka}},
  \bibinfo{journal}{Solid State Comm.} \textbf{\bibinfo{volume}{81}},
  \bibinfo{pages}{557} (\bibinfo{year}{1992}).

\bibitem[{\citenamefont{Stesliska et~al.}(2002)\citenamefont{Steslicka, Kucharczyk, Akjouj, Djafari-Rouhani, Dobrzynski and  Davison}}]{Steslicka2002}
  \bibinfo{author}{\bibfnamefont{M.}~\bibnamefont{St{\c{e}}\'slicka}},
  \bibinfo{author}{\bibfnamefont{R.}~\bibnamefont{Kucharczyk}},
  \bibinfo{author}{\bibfnamefont{A.}~\bibnamefont{Akjouj}},
  \  \bibinfo{author}{\bibfnamefont{B.}~\bibnamefont{Djafari-Rouhani}},
\bibinfo{author}{\bibfnamefont{L.}~\bibnamefont{Dobrzynski}},
 \bibnamefont{and}
  \bibinfo{author}{\bibfnamefont{S.~G..}~\bibnamefont{Davison}},
  \bibinfo{journal}{Surf. Scien. Rep.} \textbf{\bibinfo{volume}{47}},
  \bibinfo{pages}{92} (\bibinfo{year}{2002}).



\end{thebibliography}

\end{document}